\documentclass[10pt,journal,compsoc]{IEEEtran}
\ifCLASSOPTIONcompsoc
\else
\fi
\ifCLASSINFOpdf
\else
\fi
\usepackage[cmex10]{amsmath}
\usepackage{amsmath,bm}
\usepackage{algorithm}
\usepackage{algorithmic}
\usepackage[tight,footnotesize]{subfigure}
\usepackage{url}

\usepackage{graphicx}
\usepackage[noadjust]{cite}
\usepackage{color}

\begin{document}
\bstctlcite{IEEEexample:BSTcontrol}
%
\title{Community Detection via Maximization of Modularity and Its Variants}
%
%
%
%

\author{Mingming~Chen,
        Konstantin~Kuzmin, Student Member, IEEE, \\
        and~Boleslaw~K.~Szymanski, Fellow, IEEE
}

\IEEEtitleabstractindextext{%
\begin{abstract}
In this paper, we first discuss the definition of modularity ($Q$) used as a metric for community quality and then we review the modularity maximization approaches which were used for community detection in the last decade. Then, we discuss two opposite yet coexisting problems of modularity optimization: in some cases, it tends to favor small communities over large ones while in others, large communities over small ones (so called the resolution limit problem). Next, we overview several community quality metrics proposed to solve the resolution limit problem and discuss \textit{Modularity Density} ($Q_{ds}$) which simultaneously avoids the two problems of modularity. Finally, we introduce two novel fine-tuned community detection algorithms that iteratively attempt to improve the community quality measurements by splitting and merging the given network community structure. The first of them, referred to as \textit{Fine-tuned} $Q$, is based on modularity ($Q$) while the second one is based on \textit{Modularity Density} ($Q_{ds}$) and denoted as \textit{Fine-tuned} $Q_{ds}$. Then, we compare the greedy algorithm of modularity maximization (denoted as \textit{Greedy} $Q$), \textit{Fine-tuned} $Q$, and \textit{Fine-tuned} $Q_{ds}$ on four real networks, and also on the classical clique network and the LFR benchmark networks, each of which is instantiated by a wide range of parameters. The results indicate that \textit{Fine-tuned} $Q_{ds}$ is the most effective among the three algorithms discussed. Moreover, we show that \textit{Fine-tuned} $Q_{ds}$ can be applied to the communities detected by other algorithms to significantly improve their results.
\end{abstract}

}

\maketitle

\IEEEdisplaynontitleabstractindextext

%
\IEEEpeerreviewmaketitle

\section{Introduction}
%
%

%
%
%
%

\IEEEPARstart{M}{any} networks, including Internet, citation networks, transportation networks, email networks, and social and biochemical networks, display community structure which identifies groups of nodes within which connections are denser than between them \cite{UWDModularity}. Detecting and characterizing such community structure, which is known as community detection, is one of the fundamental issues in the study of network systems. Community detection has been shown to reveal latent yet meaningful structure in networks such as groups in online and contact-based social networks, functional modules in protein-protein interaction networks, groups of customers with similar interests in online retailer user networks, groups of scientists in interdisciplinary collaboration networks, etc. \cite{CommunityReport}.

In the last decade, the most popular community detection methods have been to maximize the quality metric known as modularity \cite{UWDModularity,PNASModularity,DirectedModularity,WeightedModularity} over all possible partitions of a network. Such modularity optimization algorithms include greedy algorithms \cite{NewmanGreedy,ModularityLargeNet,Wakita_Tsurumi,Louvain}, spectral methods \cite{PNASModularity,EigenvectorCommunity,Tripartition,WS,KCUT,QCUT,SpectralEquivalent}, extremal optimization \cite{ExtremalQ}, simulated annealing \cite{QMaxWithSA,QMaxWithSA2,QMaxWithSAMassen,QMaxWithSAMedus}, sampling technique \cite{SampleTechOne}, and mathematical programming \cite{MathProgrammingQ}. Modularity measures the difference between the actual fraction of edges within the community and such fraction expected in a randomized graph with the same number of nodes and the same degree sequence. It is widely used as a measurement of strength of the community structures detected by the community detection algorithms. However, modularity maximization has two opposite yet coexisting problems. In some cases, it tends to split large communities into two or more small communities \cite{QdsConference,QdsJournal}. In other cases, it tends to form large communities by merging communities that are smaller than a certain threshold which depends on the total number of edges in the network and on the degree of inter-connectivity between the communities. The latter problem is also known as the \textit{resolution limit problem} \cite{QdsConference,QdsJournal,ResolutionLimit}.

To solve these two issues of modularity, several community quality metrics were introduced, including \textit{Modularity Density} ($Q_{ds}$) \cite{QdsConference,QdsJournal} which simultaneously avoids both of them. We then propose two novel fine-tuned community detection algorithms that repeatedly attempt to improve the quality measurements by splitting and merging the given community structure. We denote the corresponding algorithm based on modularity ($Q$) as \textit{Fine-tuned} $Q$ while the one based on \textit{Modularity Density} ($Q_{ds}$) is referred to as \textit{Fine-tuned} $Q_{ds}$. Finally, we evaluate the greedy algorithm of modularity maximization (denoted as \textit{Greedy} $Q$), \textit{Fine-tuned} $Q$, and \textit{Fine-tuned} $Q_{ds}$ by using seven community quality metrics based on ground truth communities. These evaluations are conducted on four real networks, and also on the classical clique network and the LFR benchmark networks, each of which is instantiated by a wide range of parameters. The results indicate that \textit{Fine-tuned} $Q_{ds}$ is the most effective method and can also dramatically improve the community detection results of other algorithms. Further, all seven quality measurements based on ground truth communities are consistent with $Q_{ds}$, but not consistent with $Q$, which implies the superiority of \textit{Modularity Density} over the original modularity.

\section{Review of Modularity Related Literature}
In this section, we first review the definition of modularity and the corresponding optimization approaches. Then, we discuss the two opposite yet coexisting problems of modularity maximization. Finally, we overview several community quality measurements proposed to solve the resolution limit problem and then discuss \textit{Modularity Density} ($Q_{ds}$) \cite{QdsConference,QdsJournal} which simultaneously avoids these two problems.

\subsection{Definition of Modularity}
Comparing results of different network partitioning algorithms can be challenging, especially when network structure is not known beforehand. A concept of modularity defined in \cite{UWDModularity} provides a measure of the quality of a particular partitioning of a network. Modularity ($Q$) quantifies the community strength by comparing the fraction of edges within the community with such fraction when random connections between the nodes are made. The justification is that a community should have more links between themselves than a random gathering of people. Thus, the $Q$ value close to 0 means that the fraction of edges inside communities is no better than the random case, and the value of 1 means that a network community structure has the highest possible strength.

Formally, modularity ($Q$) can be defined as \cite{UWDModularity}:

\begin{equation}
\label{eq:Q}
Q= \sum\limits_{c_i \in C} \left [ \frac{\left | E_{c_i}^{in}\right |}{\left | E \right |} -\left ( \frac{2 \left | E_{c_i}^{in}\right | + \left | E_{c_i}^{out}\right |}{2 \left | E \right | } \right ) ^2 \right ],
\end{equation}
where $C$ is the set of all the communities, $c_i$ is a specific community in $C$, $\left | E_{c_i}^{in}\right |$ is the number of edges between nodes within community $c_i$, $\left | E_{c_i}^{out}\right |$ is the number of edges from the nodes in community $c_i$ to the nodes outside $c_i$, and $\vert E \vert$ is the total number of edges in the network.

Modularity can also be expressed in the following form \cite{PNASModularity}:
\begin{equation}
Q=\frac{1}{2|E|} \sum\limits_{ij} \left[A_{ij} - \frac{k_i k_j}{2|E|}\right] \delta_{{c_i}, {c_j}},
\end{equation}
where $k_i$ is the degree of node $i$, $A_{ij}$ is an element of the adjacency matrix, $\delta_{{c_i}, {c_j}}$ is the Kronecker delta symbol, and $c_i$ is the label of the community to which node $i$ is assigned.

Since larger $Q$ means a stronger community structure, several algorithms which we will discuss in the next section, are based on modularity optimization.

The modularity measure defined above is suitable only for undirected and unweighted networks. However, this definition can be naturally extended to apply to directed networks as well as to weighted networks. Weighted and directed networks contain more information than undirected and unweighted ones and are therefore often viewed as more valuable but also as more difficult to analyze than their simpler counterparts.

The revised definition of modularity that works for directed networks is as follows \cite{DirectedModularity}:
\begin{equation}
Q=\frac{1}{\vert E \vert} \sum\limits_{ij} \left[A_{ij} - \frac{k_i^{in} k_j^{out}}{|E|}\right] \delta_{{c_i}, {c_j}},
\end{equation}
where $k_i^{in}$ and $k_j^{out}$ are the in- and out- degrees.

Although many networks can be regarded as binary, i.e. as either having an edge between a pair of nodes or not having it, there are many other networks for which it would be natural to treat edges as having a certain degree of strength or weight.

The same general techniques that have been developed for unweighted networks are applied to its weighted counterparts in \cite{WeightedModularity} by mapping weighted networks onto multigraphs.
For non-negative integer weights, an edge with weight $w$ in a weighted graph corresponds to $w$ parallel edges in a corresponding multigraph. Although negative weights can arise in some applications they are rarely useful in social networks, so for the sake of brevity we will not discuss them here.
It turns out that an adjacency matrix of a weighted graph is equivalent to that of a multigraph with unweighted edges. Since the structure of adjacency matrix is independent of the edge weights, it is possible to adjust all the methods developed for unweighted networks to the weighted ones.

It is necessary to point out that the notion of degree of a node should also be extended for the weighted graphs. In this case degree of a node is defined as the sum of weights of all edges incident to this node.

It is shown in \cite{WeightedModularity} that the same definitions of modularity that were given above hold for the weighted networks as well if we treat $A_{ij}$ as the value that represents weight of the connection and set $|E|=\frac{1}{2} \sum\limits_{ij} A_{ij}$.

\subsection{Modularity Optimization Approaches}
In the literature, a high value of modularity ($Q$) indicates a good community structure and the partition corresponding to the maximum value of modularity on a given graph is supposed to have the highest quality, or at least a very good one. Therefore, it is natural to discover communities by maximizing modularity over all possible partitions of a network. However, it is computationally prohibitively expensive to exhaustively search all such partitions for the optimal value of modularity since modularity optimization is known to be NP-hard \cite{Modularity_NP}. However, many heuristic methods were introduced to find high-modularity partitions in a reasonable time. Those approaches include greedy algorithms \cite{NewmanGreedy,ModularityLargeNet,Wakita_Tsurumi,Louvain}, spectral methods \cite{PNASModularity,EigenvectorCommunity,Tripartition,WS,KCUT,QCUT,SpectralEquivalent}, extremal optimization \cite{ExtremalQ}, simulated annealing \cite{QMaxWithSA,QMaxWithSA2,QMaxWithSAMassen,QMaxWithSAMedus}, sampling technique \cite{SampleTechOne}, and mathematical programming \cite{MathProgrammingQ}. In this section, we will review those modularity optimization heuristics.

\subsubsection{Greedy Algorithms}
The first greedy algorithm was proposed by Newman \cite{NewmanGreedy}. It is a agglomerative hierarchical clustering method. Initially, every node belongs to its own community, creating altogether $|V|$ communities. Then, at each step, the algorithm repeatedly merges pairs of communities together and chooses the merger for which the resulting modularity is the largest. The change in $Q$ upon joining two communities $c_i$ and $c_j$ is
\begin{equation}
\Delta Q_{c_i,c_j}=2\left(\frac{|E_{c_i,c_j}|}{2|E|}-\frac{|E_{c_i}||E_{c_j}|}{4|E|^2}\right),
\end{equation}
where $|E_{c_i,c_j}|$ is the number of edges from community $c_i$ to community $c_j$ and $|E_{c_i}|=2|E_{c_i}^{in}|+|E_{c_i}^{out}|$ is the total degrees of nodes in community $c_i$. $\Delta Q_{c_i,c_j}$ can be calculated in constant time. The algorithm stops when all the nodes in the network are in a single community after $(|V|-1)$ steps of merging. Then, there are totally $|V|$ partitions, the first one defined by the initial step and each subsequent one resulting from each of the subsequent  $(|V|-1)$ merging steps. The partition with the largest value of modularity, approximating the modularity maximum best, is the result of the algorithm. At each merging step, the algorithm needs to compute the change $\Delta Q_{c_i,c_j}$ of modularity resulting from joining any two currently existing communities $c_i$ and $c_j$ in order to choose the best merger. Since merging two disconnected communities will not increase the value of modularity, the algorithm checks only the merging of connected pairs of communities and the number of such pairs is at most $|E|$ limiting the complexity of this part to $O(|E|)$. However, the rows and columns of adjacent matrix corresponding to the two merged communities must be updated, which takes $O(|V|)$. Since there are $(|V|-1)$ iterations, the final complexity of the algorithm is $O((|E|+|V|)|V|)$, or $O(|V|^2)$ for sparse networks.

Although Newman's algorithm \cite{NewmanGreedy} is much faster than the algorithm of Newman and Girvan \cite{UWDModularity} whose complexity is $O(|E|^2|V|)$, Clauset et al. \cite{ModularityLargeNet} pointed out that the update of the adjacent matrix at each step contains a large number of unnecessary operations when the network is sparse and therefore its matrix has a lot of zero entries. They introduced data structures for sparse matrices to perform the updating operation more efficiently. In their algorithm, instead of maintaining the adjacent matrix and computing $\Delta Q_{c_i,c_j}$, they maintained and updated the matrix with entries being $\Delta Q_{c_i,c_j}$ for the pairs of connected communities $c_i$ and $c_j$. The authors introduced three data structures to represent sparse matrices efficiently: (1) each row of the matrix is stored as a balanced binary tree in order to search and insert elements in $O(log|V|)$ time and also as a max-heap so as to locate the largest element of each row in constant time; (2) another max-heap stores the largest element of each row of the matrix so as to locate the largest $\Delta Q_{c_i,c_j}$ in constant time; (3) a vector is used to save $|E_{c_i}|$ for each community $c_i$. Then, in each step, the largest $\Delta Q_{c_i,c_j}$ can be found in constant time and the update of the adjacent matrix after merging two communities $c_i$ and $c_j$ takes $O((k_{c_i}+k_{c_j})log|V|)$, where $k_{c_i}$ and $k_{c_j}$ are the numbers of neighboring communities of communities $c_i$ and $c_j$, respectively. Thus, the total running time is at most $O(log|V|)$ times the sum of the degrees of nodes in the communities along the dendrogram created by merging steps. This sum is in the worst case the depth of the dendrogram times the sum of the degrees of nodes in the network. Suppose the dendrogram has depth $d$, then the running time is $O(d|E|log|V|)$, or $O(|V|log^2|V|)$ when the network is sparse and the dendrogram is almost balanced ($d \sim log|V|$).

However, Wakita and Tsurumi \cite{Wakita_Tsurumi} observed that the greedy algorithm proposed by Clauset et al. is not scalable to networks with sizes larger than $500,000$ nodes. They found that the computational inefficiency arises from merging communities in an unbalanced manner, which yields very unbalanced dendrograms. In such cases, the relation $d \sim log|V|$ does not hold any more, causing the algorithm to run at its worst-case complexity. To balance the merging of communities, the authors introduced three types of \textit{consolidation ratios} to measure the balance of the community pairs and used it with modularity to perform the joining process of communities without bias. This modification enables the algorithm to scale to networks with sizes up to $10,000,000$. It also approximates the modularity maximum better than the original algorithm.

Another type of greedy modularity optimization algorithm different from those above was proposed by Blondel et al., and it is usually referred to as Louvain \cite{Louvain}. It is divided into two phases that are repeated iteratively. Initially, every node belongs to the community of itself, so there are $|V|$ communities. In this first phase, every node, in a certain order, is considered for merging into its neighboring communities and the merger with the largest positive gain is selected. If all possible gains associated with the merging of this node are negative, then it stays in its original community. This merging procedure repeats iteratively and stops when no increase of $Q$ can be achieved.

After the first phase, Louvain reaches a local maximum of $Q$. Then, the second phase of Louvain builds a community network based on the communities discovered in the first phase. The nodes in the new network are the communities from the first phase and there is a edge between two new nodes if there are edges between nodes in the corresponding two communities. The weights of those edges are the sum of the weights of the edges between nodes in the corresponding two communities. The edges between nodes of the same community of the first phase result in a self-loop for this community node in the new network. After the community network is generated, the algorithm applies the first phase again on this new network. The two phases repeat iteratively and stop when there is no more change and consequently a maximum modularity is obtained. The number of iterations of this algorithm is usually very small and most of computational time is spent in the first iteration. Thus, the complexity of the algorithm grows like $O(|E|)$. Consequently, it is scalable to large networks with the number of nodes up to a billion. However, the results of Louvain are impacted by the order in which the nodes in the first phase are considered for merging \cite{ConstantCommunity}.

\subsubsection{Spectral Methods}
\label{spectral_method}
There are two categories of spectral algorithms for maximizing modularity: one is based on the modularity matrix \cite{PNASModularity,EigenvectorCommunity,Tripartition}; the other is based on the Laplacian matrix of a network \cite{WS,KCUT,QCUT}.

\textbf{\textit{A}. Modularity optimization using the eigenvalues and eigenvectors of the modularity matrix \cite{PNASModularity,EigenvectorCommunity,Tripartition}.}

Modularity ($Q$) can be expressed as \cite{PNASModularity}
\begin{equation}
\label{eq:Qmatrix}
\begin{split}
Q&=\frac{1}{4|E|} \sum_{ij}\left(A_{ij}-\frac{k_ik_j}{2|E|}\right)(s_is_j+1) \\
&=\frac{1}{4|E|} \sum_{ij}\left(A_{ij}-\frac{k_ik_j}{2|E|}\right)s_is_j=\frac{1}{4|E|}\bm{s}^T \bm{B} \bm{s},
\end{split}
\end{equation}
where $A_{ij}$ are the elements of adjacent matrix $A$ and $\bm{s}$ is the column vector representing any division of the network into two groups. Its elements are defined as $s_i=+1$ if node $i$ belongs to the first group and $s_i=-1$ if it belongs to the second group. $\bm{B}$ is the modularity matrix with elements
\begin{equation}
\label{eq:modularityMatrix}
B_{ij}=A_{ij}-\frac{k_ik_j}{2|E|}.
\end{equation}
Representing $\bm{s}$ as a linear combination of the normalized eigenvectors $\bm{u_i}$ of $\bm{B}$: $\bm{s}=\sum_{i=1}^{|V|}a_i\bm{u_i}$ with $a_i=\bm{u_i}^T\cdot\bm{s}$, and then plugging the result into Equation~(\ref{eq:Qmatrix}) yield
\begin{equation}
Q=\frac{1}{4|E|}\sum_{i}a_i\bm{u_i}^T\bm{B}\sum_ja_j\bm{u_j}=\frac{1}{4|E|}\sum_{i}a_i^2\beta_i,
\end{equation}
where $\beta_i$ is the eigenvalue of $\bm{B}$ corresponding to eigenvector $\bm{u_i}$. To maximize $Q$ above, Newman \cite{PNASModularity} proposed a spectral approach to choose $\bm{s}$ proportional to the leading eigenvector $\bm{u_1}$ corresponding to the largest (most positive) eigenvalue $\beta_1$. The choice assumes that the eigenvalues are labeled in decreasing order $\beta_1 \ge \beta_2 \ge ... \ge \beta_{|V|}$. Nodes are then divided into two communities according to the signs of the elements in $\bm{s}$ with nodes corresponding to positive elements in $\bm{s}$ assigned to one group and all remaining nodes to another. Since the row and column sums of $\bm{B}$ are zero, it always has an eigenvector $(1, 1, 1, ...)$ with eigenvalue zero. Therefore, if it has no positive eigenvalue, then the leading eigenvector is $(1, 1, 1, ...)$, which means that the network is indivisible. Moreover, Newman \cite{PNASModularity} proposed to divide network into more than two communities by repeatedly dividing each of the communities obtained so far into two until the additional contribution $\Delta Q$ to the modularity made by the subdivision of a community $c$
\begin{equation}
\begin{split}
\Delta Q&=\frac{1}{2|E|}\left[\frac{1}{2}\sum_{i,j \in c}B_{ij}(s_is_j+1)-\sum_{i,j \in c}B_{ij}\right]\\
&=\frac{1}{4|E|}\bm{s}^T\bm{B}^{(c)}\bm{s}
\end{split}
\end{equation}
is equal to or less than $0$. $\bm{B^{(c)}}$ in the formula above is the generalized modularity matrix. Its elements, indexed by the labels $i$ and $j$ of nodes within community $c$, are
\begin{equation}
B_{ij}^{(c)}=B_{ij}-\delta_{ij}\sum_{k \in c}B_{ik}.
\end{equation}
Then, the same spectral method can be applied to $\bm{B}^{(c)}$ to maximize $\Delta Q$. The recursive subdivision process stops when $\Delta Q \le 0$, which means that there is no positive eigenvalue of the matrix $\bm{B}^{(c)}$. The overall complexity of this algorithm is $O((|E|+|V|)|V|)$.

However, the spectral algorithm described above has two drawbacks. First, it divides a network into more than two communities by repeated division instead of getting all the communities directly in a single step. Second, it only uses the leading eigenvector of the modularity matrix and ignores all the others, losing all the useful information contained in those eigenvectors. Newman later proposed to divide a network into a set of communities $C$ with $|C| \ge 2$ directly using multiple leading eigenvectors \cite{EigenvectorCommunity}. Let $\bm{S}=(\bm{s_c})$ be an $|V| \times |C|$ ``community-assignment'' matrix with one column for each community $c$ defined as
\begin{equation}
 S_{i,c}=\begin{cases}
  1& \text{if node $i$ belongs to community $c$}, \\
  0& \text{otherwise}.
  \end{cases}
\end{equation}
then the modularity ($Q$) for this direct division of the network is given by
\begin{equation}
Q=\frac{1}{2|E|}\sum_{i,j=1}^{|V|} \sum_{c \in C} B_{ij} S_{i,c} S_{j,c}=\frac{1}{2|E|}\text{Tr}(\bm{S}^T\bm{B}\bm{S}),
\end{equation}
where $\text{Tr}(\bm{S}^T\bm{B}\bm{S})$ is the trace of matrix $\bm{S}^T\bm{B}\bm{S}$. Defining $\bm{B}=\bm{U}\bm{\Sigma}\bm{U}^T$, where $\bm{U}=(\bm{u_1},\bm{u_2},...)$ is the matrix of eigenvectors of $\bm{B}$ and $\bm{\Sigma}$ is the diagonal matrix of eigenvalues $\Sigma_{ii}=\beta_i$, yields
\begin{equation}
Q=\frac{1}{2|E|}\sum_{i=1}^{|V|} \sum_{c \in C} \beta_i(\bm{u_i}^T\bm{s_c})^2.
\end{equation}
Then, obtaining $|C|$ communities is equivalent to selecting $|C|-1$ independent, mutually orthogonal columns $\bm{s_c}$. Moreover, $Q$ would be maximized by choosing the columns $\bm{s_c}$ proportional to the leading eigenvectors of $\bm{B}$. However, only the eigenvectors corresponding to the positive eigenvalues will contribute positively to the modularity. Thus, the number of positive eigenvalues, plus $1$, is the upper bound of $|C|$. More general modularity maximization is to keep the leading $p~(1 \le p \le |V|)$ eigenvectors. $Q$ can be rewritten as
\begin{equation}
\begin{split}
Q&=\frac{1}{2|E|}\left(|V|\alpha+\text{Tr}[\bm{S}^T\bm{U}(\bm{\Sigma}-\alpha\bm{I})\bm{U}^T\bm{S}]\right) \\
&=\frac{1}{2|E|}\left(|V|\alpha+\sum_{j=1}^{|V|} \sum_{c \in C}(\beta_j-\alpha)\left[\sum_{i=1}^{|V|}U_{ij}S_{i,c}\right]^2\right),
\end{split}
\end{equation}
where $\alpha~(\alpha \le \beta_p)$ is a constant related to the approximation for $Q$ obtained by only adopting the first $p$ leading eigenvectors. By selecting $|V|$ node vectors $\bm{r_i}$ of dimension $p$ whose $j$th component is
\begin{equation}
[\bm{r_i}]_j=\sqrt{\beta_j-\alpha}U_{ij},
\end{equation}
modularity can be approximated as
\begin{equation}
Q \simeq \widetilde{Q} = \frac{1}{2|E|}\left(|V|\alpha+\sum_{c \in C} |\bm{R_c}|^2\right),
\end{equation}
where $\bm{R_c}$, $c \in C$, are the community vectors
\begin{equation}
\bm{R_c}=\sum_{i \in c} \bm{r_i}.
\end{equation}
Thus, the community detection problem is equivalent to choosing such a division of nodes into $|C|$ groups that maximizes the magnitudes of the community vectors $\bm{R_c}$ while requiring that $\bm{R_c} \cdot r_i > 0$ if node $i$ is assigned to community $c$. Problems of this type are called \textit{vector partitioning} problems.

Although \cite{EigenvectorCommunity} explored using multiple leading eigenvectors of the modularity matrix, it did not pursue it in detail beyond a two-eigenvector approach for bipartitioning \cite{PNASModularity,EigenvectorCommunity}. Richardson et al. \cite{Tripartition} provided a extension of these recursive bipartitioning methods by considering the best two-way or three-way division at each recursive step to more thoroughly explore the promising partitions. To reduce the number of partitions considered for the eigenvector-pair tripartitioning, the authors adopted a divide-and-conquer method and as a result yielded an efficient approach whose computational complexity is competitive with the two-eigenvector bipartitioning method.

\textbf{\textit{B}. Modularity optimization using the eigenvalues and eigenvectors of the Laplacian matrix \cite{WS,KCUT,QCUT}.}

Given a partition $C$ (a set of communities) and the corresponding ``community-assignment'' matrix $\bm{S}=(\bm{s_c})$, White and Smyth \cite{WS} rewrote modularity ($Q$) as follows:
\begin{equation}
Q \propto \text{Tr}(\bm{S}^T(W-\widetilde{D})\bm{S})=-\text{Tr}(\bm{S}^T\bm{L_Q}\bm{S}),
\end{equation}
where $W=2|E|A$ and the elements of $\widetilde{D}$ are $\widetilde{D}_{ij}=k_ik_j$. The matrix $\bm{L_Q}=\widetilde{D}-W$ is called the ``Q-Laplacian''. Finding the ``community-assignment'' matrix $\bm{S}$ that maximizes $Q$ above is NP-complete, but a good approximation can be obtained by relaxing the discreteness constraints of the elements of $\bm{S}$ and allowing them to assume real values. Then, $Q$ becomes a continuous function of $\bm{S}$ and its extremes can be found by equating its first derivative with respect to $\bm{S}$ to zero. This leads to the eigenvalue equation:
\begin{equation}
\label{eq:QLaplacian}
\bm{L_Q}\bm{S}=\bm{S}\bm{\Lambda},
\end{equation}
where $\bm{\Lambda}$ is the diagonal matrix of Lagrangian multipliers. Thus, the modularity optimization problem is transformed into the standard spectral graph partitioning problem. When the network is not too small, $\bm{L_Q}$ can be approximated well, up to constant factors, by the transition matrix $\bm{\widetilde{W}}=D^{-1}A$ obtained by normalizing $A$ so that all rows sum to one. $D$ here is the diagonal degree matrix of $A$. It can be shown that the eigenvalues and eigenvectors of $\bm{\widetilde{W}}$ are precisely $1-\lambda$ and $\mu$, where $\lambda$ and $\mu$ are the solutions to the generalized eigenvalue problem $\bm{L}\mu=\lambda D \mu$ where $\bm{L}=D-A$ is the Laplacian matrix. Thus, the underlying spectral algorithm here is equivalent to the standard spectral graph partitioning problem which uses the eigenvalues and eigenvectors of the Laplacian matrix.

Based on the above analysis, White and Smyth proposed two clustering algorithms, named ``Algorithm Spectral-1'' and ``Algorithm Spectral-2'', to search for a partition $C$ with size up to $K$ predefined by an input parameter. Both algorithms take the eigenvector matrix $\bm{U_{K}}=(\bm{u_1}, \bm{u_2}, ..., \bm{u_{K-1}})$ with the leading $K-1$ eigenvectors (excluding the trivial all-ones eigenvector) of the transition matrix $\bm{\widetilde{W}}$ as input. Those $K-1$ eigenvectors can be efficiently computed with the Implicitly Restarted Lanczos Method (IRLM) \cite{Lanczos}. ``Algorithm Spectral-1'' uses the first $k-1$ ($2 \le k \le K$) columns of $\bm{U_{K}}$, denoted as $\bm{U_{k-1}}$, and clusters the row vectors of $\bm{U_{k-1}}$ using $k$-means to find a $k$-way partition, denoted as $C_k$. Then, the $C_{k^*}$ with size $k^*$ that achieves the largest value of $Q$ is the final community structure.

``Algorithm Spectral-2'' starts with a single community ($k=1$) and recursively splits each community $c$ into two smaller ones if the subdivision produces a higher value of $Q$. The split is done by running $k$-means with two clusters on the matrix $\bm{U_{k,c}}$ formed from $\bm{U_k}$ by keeping only rows corresponding to nodes in $c$. The recursive procedure stops when no more splits are possible or when $k=K$ communities have been found and then the final community structure with the highest value of $Q$ is the detection result.

However, the two algorithms described above, especially ``Algorithm Spectral-1'', scale poorly to large networks because of running $k$-means partitioning up to $K$ times. Both approaches have a worst-case complexity $O(K^2|V|+K|E|)$. In order to speed up the calculation while retaining effectiveness in approximating the maximum of $Q$, Ruan and Zhang \cite{KCUT} proposed the \textit{Kcut} algorithm which recursively partitions the network to optimize $Q$. At each recursive step, \textit{Kcut} adopts a $k$-way partition ($k=2,3,...,l$) to the subnetwork induced by the nodes and edges in each community using ``Algorithm Spectral-1'' of White and Smyth \cite{WS}. Then, it selects the $k$ that achieves the highest $Q$. Empirically, \textit{Kcut} with $l$ as small as $3$ or $4$ can significantly improve $Q$ over the standard bi-partitioning method and it also reduces the computational cost to $O((|V|+|E|)log|C|)$ for a final partition with $|C|$ communities.

Ruan and Zhang later \cite{QCUT} proposed \textit{QCUT} algorithm that combines \textit{Kcut} and local search to optimize $Q$. The \textit{QCUT} algorithm consists of two alternating stages: partitioning and refinement. In the partitioning stage, \textit{Kcut} is used to recursively partition the network until $Q$ cannot be further improved. In the refinement stage, a local search strategy repeatedly considers two operations. The first one is migration that moves a node from its current community to another one and the second one is the merge of two communities into one. Both are applied to improve $Q$ as much as possible. The partitioning stage and refinement stage are alternating until $Q$ cannot be increased further. In order to solve the resolution limit problem of modularity, the authors proposed \textit{HQUCT} which recursively applies \textit{QCUT} to divide the subnetwork, generated with the nodes and edges in each community, into subcommunities. Further, to avoid overpartitioning, they use a statistical test to determine whether a community indeed has intrinsic subcommunities.

\textbf{\textit{C}. Equivalence of two categories of spectral algorithms for maximizing modularity \cite{SpectralEquivalent}.}

Newman \cite{SpectralEquivalent} showed that with hyperellipsoid relaxation, the spectral modularity maximization method using the eigenvalues and eigenvectors of the modularity matrix can be formulated as the spectral algorithm that relies on the eigenvalues and eigenvectors of Laplacian matrix. This formulation indicates that the above two kinds of modularity optimization approaches are equivalent. Starting with Equation~(\ref{eq:Qmatrix}) for the division of a network into two groups, first the discreteness of $s_i$ is relaxed onto a hyperellipsoid with the constraint
\begin{equation}
\sum_{i} k_is_i^2=2|E|.
\end{equation}
Then, the relaxed modularity maximization problem can be easily solved by setting the first derivative of Equation~(\ref{eq:Qmatrix}) with respect to $s_i$ to zero. This leads to
\begin{equation}
\label{eq:relaxedEq}
\sum_{j}B_{ij}s_j=\lambda k_is_i,
\end{equation}
or in matrix notation
\begin{equation}
\label{eq:relaxedMatrixEq}
\bm{B}\bm{s}=\lambda D\bm{s},
\end{equation}
where $\lambda$ is the eigenvalue. Plugging Equation~(\ref{eq:relaxedEq}) into Equation~(\ref{eq:Qmatrix}) yields
\begin{equation}
Q=\frac{1}{4|E|} \sum_{ij}B_{ij}s_is_j=\frac{\lambda}{4|E|}\sum_{i}k_is_i^2=\frac{\lambda}{2}.
\end{equation}
Therefore, to achieve the highest value of $Q$, one should chose $\lambda$ to be the largest (most positive) eigenvalue of Equation~(\ref{eq:relaxedMatrixEq}). Using Equation~(\ref{eq:modularityMatrix}), Equation~(\ref{eq:relaxedEq}) can be rewritten as
\begin{equation}
\sum_{j} A_{ij}s_j=k_i(\lambda s_i + \frac{1}{2|E|}\sum_j k_js_j),
\end{equation}
or in matrix notion as
\begin{equation}
\bm{A}\bm{s}=D\left(\lambda \bm{s}+\frac{\bm{k}^T\bm{s}}{2|E|}\bm{1}\right),
\end{equation}
where $\bm{k}$ is the vector with element $k_i$ and $\bm{1}=(1,1,1,...)$. Then, multiplying the above equation by $\bm{1}^T$ results in $\lambda \bm{k}^T\bm{s}=0$. If there is a nontrivial eigenvalue $\lambda>0$, then the above equation simplifies to
\begin{equation}
\label{eq:simplifiedEq}
\bm{A}\bm{s}=\lambda D \bm{s}.
\end{equation}
Again, $\lambda$ should be the most positive eigenvalue. However, the eigenvector corresponding to this eigenvalue is the uniform vector $\bm{1}$ which fails to satisfy $\bm{k}^T\bm{s}=0$. Thus, in this case, one can do the best by choosing $\lambda$ to be the second largest eigenvalue and having $\bm{s}$ proportional to the corresponding eigenvector. In fact, this eigenvector is precisely equal to the leading eigenvector of Equation~(\ref{eq:relaxedMatrixEq}). Then, after defining a rescaled vector $\bm{u}=\bm{D}^{1/2}\bm{s}$ and plugging it into Equation~(\ref{eq:simplifiedEq}), we get
\begin{equation}
(\bm{D}^{-1/2}\bm{A}\bm{D}^{-1/2})\bm{u}=\lambda \bm{u}.
\end{equation}
The matrix $\bm{L}=\bm{D}^{-1/2}\bm{A}\bm{D}^{-1/2}$ is called the normalized Laplacian matrix. (The normalized Laplacian is sometimes defined as $\bm{L}=\bm{I}-\bm{D}^{-1/2}\bm{A}\bm{D}^{-1/2}$, but those two differ only by a trivial transformation of their eigenvalues and eigenvectors.)

\subsubsection{Extremal Optimization}
Duch and Arenas \cite{ExtremalQ} proposed a modularity optimization algorithm based on the Extremal Optimization (EO) \cite{EO}. EO optimizes a global variable by improving extremal local variables. Here, the global variable is modularity ($Q$). The contribution of an individual node $i$ to $Q$ of the whole network with a certain community structure is given by
\begin{equation}
q_i=k_{i,c}-k_i\frac{|E_c|}{2|E|},
\end{equation}
where $k_{i,c}$ is the number of edges that connect node $i$ to the nodes in its own community $c$. Notice that $Q=\frac{1}{2|E|}\sum_i q_i$ and $q_i$ can be normalized into the interval $[-1,1]$ by diving it by $k_i$
\begin{equation}
\lambda_i=\frac{q_i}{k_i}=\frac{k_{i,c}}{k_i}-\frac{|E_c|}{2|E|},
\end{equation}
where $\lambda_i$, called fitness, is the relative contribution of node $i$ to $Q$. Then, the fitness of each node is adopted as the local variable.

The algorithm starts by randomly splitting the network into two partitions of equal number of nodes, where communities are the connected components in each partition. Then, at each iteration, it moves the node with the lowest fitness from its own community to another community. The shift changes the community structure, so the fitness of many other nodes needs to be recomputed. The process repeats until it cannot increase $Q$. After that, it generates sub-community networks by deleting the inter-community edges and proceeds recursively on each sub-community network until $Q$ cannot be improved. Although the procedure is deterministic when given the initialization, its final result in fact depends on the initialization and it is likely to get trapped in local maxima. Thus, a probabilistic selection called $\tau$-EO \cite{EO} in which nodes are ranked according to their fitness and a node of rank $r$ is selected with the probability $P(r) \propto r^{-\tau}$ is used to improve the result. The computational complexity of this algorithm is $O(|V|^2log^2|V|)$.

\subsubsection{Simulated Annealing}
Simulated annealing (SA) \cite{SA} is a probabilistic procedure for the global optimization problem of locating a good approximation to the global optimum of a given function in a large search space. This technique was adopted in \cite{QMaxWithSA,QMaxWithSA2,QMaxWithSAMassen,QMaxWithSAMedus} to maximize modularity ($Q$). The initial point for all those approaches can be arbitrary partitioning of nodes into communities, even including $|V|$ communities in which each node belongs to its own community. At each iteration, a node $i$ and a community $c$ are chosen randomly. This community could be a currently existing community or an empty community introduced to increase the number of communities. Then, node $i$ is moved from its original community to this new community $c$, which would change $Q$ by $\Delta Q$. If $\Delta Q$ is greater than zero, this update is accepted, otherwise it is accepted with probability $e^{\beta \Delta Q}$ where $\beta$ in \cite{QMaxWithSA,QMaxWithSA2,QMaxWithSAMassen} represents the inverse of temperature $T$ and $\beta$ in \cite{QMaxWithSAMedus} is the reciprocal of pseudo temperature $\tau$. In addition in \cite{QMaxWithSAMedus}, there is one more condition for the move of a node when $c$ is not empty, shifting node $i$ to $c$ is considered only if there are some edges between node $i$ and the nodes in $c$. To improve the performance and to avoid getting trapped in local minima, collective movements which involve moving multiple nodes at a time \cite{QMaxWithSAMassen,QMaxWithSAMedus}, merging two communities \cite{QMaxWithSA,QMaxWithSA2,QMaxWithSAMassen}, and splitting a community \cite{QMaxWithSA,QMaxWithSA2,QMaxWithSAMassen} are employed. Splits can be carried out in a number of different schemes. The best performance is achieved by treating a community as an isolated subnetwork and partitioning it into two and then performing a nested SA on these partitions \cite{QMaxWithSA,QMaxWithSA2}. Those methods stop when no new update is accepted within a fixed number of iterations.

\subsubsection{Sampling Techniques}
Sales-Pardo et al. \cite{SampleTechOne} proposed a ``box-clustering'' method to extract the hierarchical organization of networks. This approach consists of two steps: (1) estimating the similarity, called ``node affinity'', between nodes and forming the node affinity matrix; (2) deriving hierarchical community structure from the affinity matrix. The affinity between two nodes is the probability that they are classified into the same community in the local maxima partitions of modularity. The set of local maxima partitions, called $P_{max}$, includes those partitions for which neither the moving of a node from its original community to another, nor the merging of two communities will increase the value of modularity. The sample $P_{max}$ is found by performing the simulated annealing based modularity optimization algorithm of Guimer\'{a} and Amaral \cite{QMaxWithSA,QMaxWithSA2}. More specifically, the algorithm first randomly divides the nodes into communities and then performs the hill-climbing search until a sample with local maximum of modularity is reached. Then, the affinity matrix is updated based on the obtained sample.

The sample generation procedure is repeated until the affinity matrix has converged to its asymptotic value. Empirically, the total number of samples needed is proportional to the size of the network. Before proceeding to the second step, the algorithm assesses whether the network has a significant community structure or not. It is done by computing the $z$-score of the average modularity of the partitions in $P_{max}$ with respect to the average modularity of the partitions with the local modularity maxima of the equivalent ensemble of null model networks. The equivalent null model is obtained by randomly rewiring the edges of the original network while retaining the degree sequence. Large $z$-score indicates that the network has a meaningful internal community structure. If the network indeed has a significant community structure, the algorithm advances to the second step to group nodes with large affinity close to each other. The goal is to bring the form of the affinity matrix as close as possible to block-diagonal structure by minimizing the cost function representing the average distance of matrix elements to the diagonal. Then, the communities corresponds to the ``best'' set of boxes obtained by least-squares fitting of the block-diagonal structure to the affinity matrix. The procedure described above can be recursively performed to subnetworks induced by communities to identify the low level structure of each community until no subnetwork is found to have significant intrinsic structure.

\subsubsection{Mathematical Programming}
Agarwal and Kempe \cite{MathProgrammingQ} formulated the modularity maximization problem as a linear program and vector program which have the advantage of providing a posteriori performance guarantees. First, modularity maximization can be transformed into the integer program
\begin{equation}
\begin{split}
&\text{Maximize}~\frac{1}{2|E|}\sum_{ij}B_{ij}(1-x_{ij}) \\
&\text{subject to}~x_{ik} \le x_{ij}+x_{jk}~\text{for all}~i,j,k \\
&~~~~~~~~~~~~~x_{ij} \in \{0,1\}~\text{for all}~i,j,
\end{split}
\end{equation}
where $\bm{B}$ is the modularity matrix and the objective function is linear in the variable $x_{ij}$. When $x_{ij}=0$, $i$ and $j$ belong to the same community and $x_{ij}=1$ indicates that they are in different communities. The restriction $x_{ik} \le x_{ij}+x_{jk}$ requires that $i$ and $k$ are in the same community if and only if $i$, $j$, and $k$ are in the same community. Solving the above integer program is NP-hard, but relaxing the last constraint that $x_{ij}$ is a integer from $\{0,1\}$ to allow $x_{ij}$ be a real number in the interval $[0,1]$ reduces the integer program to a linear program which can be solved in polynomial time \cite{LinearProgram}. However, the solution does not correspond to a partition when any of $x_{ij}$ is fractional. To get the communities from $x_{ij}$, a rounding step is needed. The value of $x_{ij}$ is treated as the distance between $i$ and $j$ and these distances are used repeatedly to form communities of ``nearby'' nodes. Moreover, optimizing modularity by dividing a network into two communities can be considered as a strict quadratic program
\begin{equation}
\begin{split}
&\text{Maximize}~\frac{1}{4|E|}\sum_{ij}B_{ij}(1+s_is_j) \\
&\text{subject to}~s_i^2=1~\text{for all}~i,
\end{split}
\end{equation}
where the objective function is the same as Equation~(\ref{eq:Qmatrix}) defined by Newman \cite{PNASModularity}. Note that the constraint $s_i^2=1$ ensures that $s_i=\pm 1$ which implies that node $i$ belongs either to the first or the second community. Quadratic programming is NP-complete, but it could be relaxed to a vector program by replacing each variable $s_i$ with $|V|$-dimensional vector $\bm{s}$ and replacing the scalar product with the inner vector product. The solution to vector program is one location per node on the surface of a $|V|$-dimensional hypersphere. To obtain a bipartition from these node locations, a rounding step is needed which chooses any random $(|V|-1)$-dimensional hyperplane passing through the origin and uses this hyperplane to cut the hypersphere into two halves and as a result separate the node vectors into two parts. Multiple random hyperplanes can be chosen and the one that gets the community structure with the highest modularity provides a solution. The same vector program is then recursively applied to subnetworks generated with nodes and edges in discovered communities to get hierarchical communities until $Q$ cannot be increased. Following the linear program and vector program, Agarwal and Kempe also adopted a post-processing step similar to the local search strategy proposed by Newman \cite{PNASModularity} to further improve the results.

\subsection{Resolution limit}
Since its inception, the modularity has been used extensively as the measure of the quality of partitions produced by community detection algorithms. In fact, if we adopt modularity as a quality measure of communities, the task of discovering communities is essentially turned into the task of finding the network partitioning with an optimal value of modularity.

However as properties of the modularity were studied, it was discovered that in some cases it fails to detect small communities. There is a certain threshold  \cite{ResolutionLimit}, such that a community of the size below it will not be detected even if it is a complete subgraph connected to the rest of the graph with a single edge. This property of modularity has become known as the \textit{resolution limit}.

Although the resolution limit prevents detection of small communities, the actual value of the threshold depends on the total number of edges in the network and on the degree of interconnectedness between communities. In fact, the resolution limit can reach the values comparable to the size of the entire network causing formation of a few giant communities (or even a single community) and failing to detect smaller communities within them. It makes interpreting the results of community detection very difficult because it is impossible to tell beforehand whether a community is well-formed or if it can be further split into subcommunities.

Considering modularity as a function of the total number of edges, $\vert E \vert$, and the number of communities, $m$, makes it possible to find the values of $m$ and $\vert E \vert$ which maximize this function. It turns out that setting $m = \sqrt{\vert E \vert}$ yields the absolute maximal value of modularity. Consequently, modularity has a resolution limit of order $\sqrt{\vert E \vert}$ which bounds the number and size of communities \cite{ResolutionLimit}. In fact, if for a certain community the number of edges inside it is smaller than $\sqrt{\frac{\vert E \vert}{2}}$, such community cannot be resolved through the modularity optimization. It is also possible for modularity optimization to fail to detect communities of larger size if they have more edges in common with the rest of the network. Therefore, by finding the optimal value of the modularity we are generally not obtaining the best possible structure of communities.

The above arguments can also be applied to weighted networks. In this case, $\vert E \vert$ is the sum of the weights of all the edges in the network, $\vert E_{c_i}^{in} \vert$ is the sum of the weights of the edges between nodes within community $c_i$, and $\vert E_{c_i}^{out} \vert$ is the sum of the weights of the edges from the nodes in community $c_i$ to the nodes outside $c_i$.

By introducing an additional parameter, $\epsilon$, which represents the weight of inter-community edges, Berry et al. showed in \cite{ResolutionLimitWeight} that the number of communities in the optimal solution is
\begin{equation}
m = \sqrt{\frac{\vert E \vert}{\epsilon}}.
\end{equation}
Correspondingly, any community for which its size
\begin{equation}
\vert c_i \vert < \sqrt{\frac{\vert E \vert \epsilon}{2}} - \epsilon
\end{equation}
 may not be resolved.

Introduction of $\epsilon$ brings some interesting opportunities. If we can make $\epsilon$ arbitrarily small, then we can expect maximum weighted modularity to produce any desired number of communities. In other words, given a proper weighting, a much better modularity resolution can be achieved than without weighting. However, in practice, finding a way to set edge weights to achieve small values of $\epsilon$ can be challenging. An algorithm for lowering $\epsilon$ proposed by Berry et al. requires $O(m|V| \log |V|)$ time.


\subsection{Resolving the resolution limit problem}
There have been extensive studies done on how to mitigate the consequences of the modularity resolution limit. The main approaches followed are described below.

Localized modularity measure ($LQ$) \cite{muff2005local} is based on the observation that the resolution limit problem is caused by modularity being a global measure since it assumes that edges between any pairs of nodes are equally likely, including connectivity between the communities. However, in many networks, the majority of communities have edges to only a few other communities, i.e. exhibit a local community connectivity.

Thus, a local version of the modularity measure for a directed network is defined as:
\begin{equation}
LQ= \sum\limits_{c_i \in C} \left [ \frac{\left | E_{c_i}^{in}\right |}{\left | E_{c_i}^{neighb} \right |} -\left (\frac{\left | E_{c_i}^{in}\right | + \left | E_{c_i}^{out}\right |}{ \left | E_{c_i}^{neighb} \right | } \right )^2 \right ],
\end{equation}
where $\left | E_{c_i}^{neighb} \right |$ is the total number of edges in the neighboring communities of $c_i$, i.e. in the communities to which all neighbors of $c_i$ belong.

Unlike traditional modularity ($Q$), the local version of modularity ($LQ$) is not bounded above by 1. The more locally connected communities a network has, the bigger its $LQ$ can grow. In a network where all communities are connected to each other, $LQ$ yields the same value as $Q$. $LQ$ considers individual communities and their neighbors, and therefore provides a measure of community quality that is not dependent on other parts of the network. The local connectivity approach can be applied not only to the nearest neighboring communities, but also to the second or higher neighbors as well.

Arenas et al. proposed a multiple resolution method \cite{arenas2008analysis} which is based on the idea that it might be possible to look at the detected community structure at different scales. From this perspective, the modularity resolution limit is not a problem but a feature. It allows choosing a desired resolution level to achieve the required granularity of the output community structure using the original definition of modularity.

The multiple resolution method is based on the definition of modularity given by Equation~(\ref{eq:Q}). The modularity resolution limit depends on the total weight $2 \left | E \right |$. By varying the total weight, it is possible to control the resolution limit, effectively performing community detection at different granularity levels. Changing the sum of weights of edges adjacent to every node by some value $r$ results in rescaling topology by a factor of $r$. Since the resolution limit is proportional to $\sqrt{r}$, the growth of the resolution limit is slower than that of $r$. Consequently, it would be possible to achieve a scale at which all required communities would be visible to the modularity optimization problem.

Caution should be exercised when altering the weights of edges in the network to avoid changing its topological characteristics. To ensure this, a rescaled adjacency matrix can be defined as:
\begin{equation}
A_r = A + rI,
\end{equation}
where $A$ is the adjacency matrix and $I$ is the identity matrix. Since the original edge weights are not altered, $A_r$ preserves all common features of the network: distribution of sum of weights, weighted clustering coefficient, eigenvectors, etc. Essentially, introducing $r$ results in a self-loop of weight $r$ being added to every node in the network.

Optimizing the modularity for the rescaled topology $A_r$ is performed by using the modularity at scale $r$ as the new quality function:
\begin{equation}
Q_r= \sum\limits_{c_i \in C} \left [ \frac{2 \left | E_{c_i}^{in}\right | + r \left | c_i \right |}{2 \left | E \right | + r \left | V \right |} -
\left ( \frac{|E_{c_i}| + r \left | c_i \right |}{ 2 \left | E \right | + r \left | V \right | } \right )^2 \right ],
\end{equation}
where $\left |c_i \right |$ is the number of nodes in community $c_i$ and $|E_{c_i}|=2|E_{c_i}^{in}|+|E_{c_i}^{out}|$.
It yields larger communities for smaller values of $r$ and smaller communities for larger values of $r$. By performing modularity optimization for different values of $r$, it is possible to analyze the community structure at different scales.

Parameter $r$ can also be thought of as representing resistance of a node to become part of a community. If $r$ is positive, we can obtain a network community structure that is more granular than what was possible to achieve with the original definition of modularity ($Q$) which corresponds to $r$ being zero. Making $r$ negative zooms out of the network and provides a view of super communities.

Further studies of the multiple resolution approach revealed that it suffers from two major issues outlined in \cite{lancichinetti2011limits}. First, when the value of the resolution parameter $r$ is low it tends to group together small communities. Second, when the resolution is high, it splits large communities. These trends are opposite for networks with a large variation of community sizes. Hence, it is impossible to select a value of the resolution parameter such that neither smaller nor larger communities are adversely affected by the resolution limit. A network can be tested for susceptibility to the resolution problem by examining its clustering coefficient, i.e. a degree to which nodes tend to form communities. If the clustering coefficient has sharp changes, it indicates that communities of substantially different scales exist in this network. The result is that when the value of $r$ is sufficiently large, bigger communities get broken up before smaller communities are found. This applies also to other multiple resolution methods and seems to be a general problem of the methods that are trying to optimize some global measure.

The hierarchical multiresolution method proposed by Granell et al. in \cite{granell2012hierarchical} overcomes the limitations of the multiple resolution method on networks with very different scales of communities. It achieves that by introducing a new hierarchical multiresolution scheme that works even in cases of community detection near the modularity resolution limit. The main idea underlying this method is based on performing multiple resolution community detection on essential parts of the network, thus analyzing each part independently.

The method operates iteratively by first placing all nodes in a singe community. Then, it finds the minimum value of the resistance parameter $r$ which produces a community structure with the optimal value of modularity. Finally, it runs the same algorithm on each community that was found. The method terminates when no more split of communities is necessary, which usually takes just a few steps.

Another approach to leveraging the results of modularity optimization has been introduced by Chakraborty et al. in \cite{ConstantCommunity}. It is based on the observation that a simple change to the order of nodes in a network can significantly affect the community structure. However, a closer examination of the communities produced in different runs of a certain community detection algorithm reveals that for many networks the same invariant groups of nodes are consistently assigned to the same communities. Such groups of nodes are called constant communities. The percentage of constant communities varies depending on the network. Constant communities are detected by trying different node permutations while preserving the degree sequence of the nodes. For networks that have strong community structure, the constant communities detected can be adopted as a pre-processing step before performing modularity optimization. This can lead to higher modularity values and lower variability in results, thus improving the overall quality of community detection.

In the study \cite{li2008quantitative} by Li, Zhang et al., a new quantitative measure for community detection is introduced. It offers several improvements over the modularity ($Q$), including elimination of the resolution limit and ability to detect the number of communities. The new measure called modularity density ($D$) is based on the average degree of the community structure. It is given by:
\begin{equation}
D= \sum\limits_{c_i \in C} \frac{2\left | E_{c_i}^{in}\right | - \left | E_{c_i}^{out}\right |}{\left | c_i \right |}.
\end{equation}
The quality of the communities found is then described by the value of the modularity density ($D$). The larger the value of $D$, the stronger the community structure is.

The modularity density ($D$) does not divide a clique into two parts, and it can resolve most modular networks correctly. It can also detect communities of different sizes. This second property can be used to quantitatively determine the number of communities, since the maximum $D$ value is achieved when the network is supposed to correctly partitioned. Although as mentioned in \cite{li2008quantitative} finding an optimal value of modularity density ($D$) is NP-hard, it is equivalent to an objective function of the kernel $k$ means clustering problem for which efficient computational algorithms are known.

Traag et al. in \cite{traag2011narrow} introduce a rigorous definition of the \textit{resolution-limit-free} method for which considering any induced subgraph of the original graph does not cause the detected community structure to change. In other words, if there is an optimal partitioning of a network (with respect to some objective function), and for each subgraph induced by the partitioning it is also optimal, then such objective function is called resolution-limit-free. An objective function is called \textit{additive} for a certain partitioning if it is equal to the sum of the values of this objective function for each of the subgraphs induced by the partitioning.

Based on these two definitions it is proved that if an objective function is additive and there are two optimal partitions, then any combination of these partitions is also optimal. In case of a complete graph, if an objective function is resolution-limit-free, then an optimal partitioning either contains all the nodes (i.e. there is only one community which includes all nodes) or consists of communities of size 1 (i.e. each node forms a community of its own). A more general statement for arbitrary objective functions is also true: if an objective function has local weights (i.e. weights that do not change when considering subgraphs) then it is resolution-limit-free. Although the converse is not true, there is only a relatively small number of special cases when methods with non-local weights are resolution-limit-free.

The authors then analyze \textit{resolution-limit-free} within the framework of the first principle Potts model \cite{PottsModel}:
\begin{equation}
\label{H}
\mathcal{H} =-\sum\limits_{ij} \left(a_{ij}A_{ij} - b_{ij}\left(1-A_{ij}\right)\right) \delta_{{c_i}, {c_j}},
\end{equation}
where $a_{ij}$, $b_{ij} \geq 0$ are some weights. The intuition behind this formula is that a community should have more edges inside it than edges which connect it to other communities. Thus, it is necessary to reward existing links inside a community and penalize links that are missing from a community. The smaller the value of $\mathcal{H}$ is, the more desirable the community structure is. However the minimal value might not be unique.

Given the definition of $\mathcal{H}$, it is possible to describe various existing community detection methods with an appropriate choice of parameters, as well as propose alternative methods. The following models are shown to fit into $\mathcal{H}$: Reichardt and Bornholdt (RB), Arenas, Fern\'{a}ndes, and G\'{o}mez (AFG), Ronhovde and Nussinov (RN) as well as the label propagation method. RB approach with a configuration null model also covers the original definition of modularity. The authors also propose a new method called constant Potts model (CPM) by choosing $a_{ij}=w_{ij}-b_{ij}$ and $b_{ij} = \gamma$ where $w_{ij}$ is the weight of the edge between nodes $i$ and $j$, and $\gamma$ is a constant. CPM is similar to RB and RN models but is simpler and more intuitive. CPM and RN have local weights and are consequently resolution-limit-free, while RB, AFG, and modularity are not.

However, all of the above approaches are aimed at solving only the resolution limit problem. Work done by Chen et al. in \cite{QdsConference,QdsJournal} adopts a different definition of modularity density which simultaneously addresses two problems of modularity. It is done by mixing two additional components, \textit{Split Penalty} ($SP$) and the community density, into the well-known definition of modularity. Community density includes internal community density and pair-wise community density. \textit{Split Penalty} ($SP$) is the fraction of edges that connect nodes of different communities:
\begin{equation}
\label{eq:uwdsp}
SP=\sum_{c_i \in C}\left[\sum_{\substack{c_j \in C \\ c_j \ne c_i}}\frac{|E_{c_i,c_j}|}{2|E|}\right].
\end{equation}
The value of \textit{Split Penalty} is subtracted from modularity, while the value of the community density is added to modularity and \textit{Split Penalty}. Introducing \textit{Split Penalty} resolves the issue of favoring small communities. Community density eliminates the problem of favoring large communities (also known as the resolution limit problem). The \textit{Modularity Density} ($Q_{ds}$) is then given by:
\begin{equation}
\label{eq:Qds}
\begin{split}
&Q_{ds}=\sum_{c_i \in C} \biggl[\frac{|E_{c_i}^{in}|}{|E|}d_{c_i}-\left(\frac{2|E_{c_i}^{in}|+|E_{c_i}^{out}|}{2|E|}d_{c_i}\right)^2 \\
& ~~~~~~~~~~~~~-\sum_{\substack{c_j \in C \\ c_j \ne c_i}}\frac{|E_{c_i,c_j}|}{2|E|}d_{c_i,c_j}\biggr], \\
& d_{c_i}=\frac{2|E_{c_i}^{in}|}{|c_i|(|c_i|-1)}, \\
& d_{c_i,c_j}=\frac{|E_{c_i,c_j}|}{|c_i||c_j|}.
\end{split}
\end{equation}
where $d_{c_i}$ is the internal density of community $c_i$, $d_{c_i,c_j}$ is the pair-wise density between community $c_i$ and community $c_j$.

\textit{Modularity Density} ($Q_{ds}$) avoids falling into the trap of merging two or more consecutive cliques in the ring of cliques network or dividing  a clique into two or more parts. It can also discover communities of different sizes. Thus, using $Q_{ds}$ solves both the resolution limit problem of modularity and the problem of splitting larger communities into smaller ones. Hence, $Q_{ds}$ is an very effective alternative to $Q$.

\section{Fine-tuned Algorithm}
In our previous papers \cite{QdsConference,QdsJournal}, we have given the definition of \textit{Modularity Density} ($Q_{ds}$). With formal proofs and experiments on two real dynamic datasets (Senate dataset \cite{SenateDataset} and Reality Mining Bluetooth Scan data \cite{RealityMining}) we demonstrated that $Q_{ds}$ solves the two opposite yet coexisting problems of modularity: the problem of favoring small communities and the problem of favoring large communities (also called the resolution limit problem). Moreover, for a given community in $Q_{ds}$ defined by Equation~(\ref{eq:Qds}), its internal and pair-wise densities and its split penalty are local components, which is related to the \textit{resolution-limit-free} definition in \cite{traag2011narrow}. Therefore, it is reasonable to expect that maximizing $Q_{ds}$ would discover more meaningful community structure than maximizing $Q$. In this section, we first illustrate why the greedy agglomerative algorithm for increasing $Q_{ds}$ cannot be adopted for optimizing $Q_{ds}$. Then, we propose a fine-tuned community detection algorithm that repeatedly attempts to improve the community quality measurements by splitting and merging the given network community structure to maximize $Q_{ds}$.

\begin{figure}[!t]
\centering
\setlength{\belowcaptionskip}{-1em}
\includegraphics[scale=0.69]{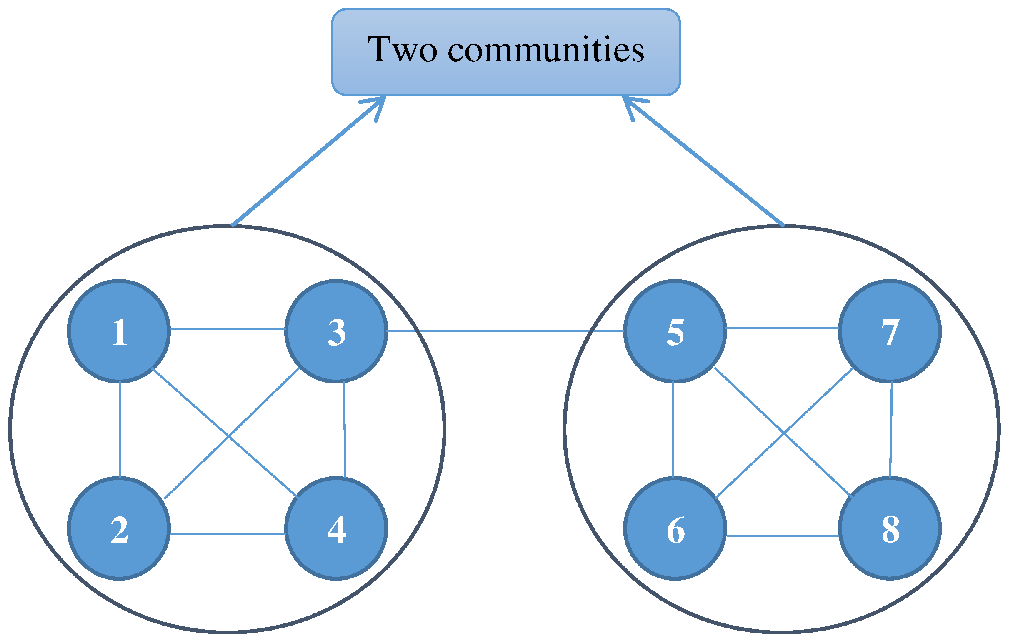}
\vspace{-0.7em}
\centering
\caption{A simple network with two clique communities. Each clique has four nodes and the two clique communities are connected to each other with one single edge.}
\label{two_clique_communities}
\vspace{-0.5em}
\end{figure}

\subsection{Greedy Algorithm Fails to Optimize $\bm{Q_{ds}}$}

In this subsection, we show why the greedy agglomerative algorithm increasing $Q_{ds}$ fails to optimize it. At the first step of the greedy  algorithm for increasing $Q_{ds}$, each node is treated as a single community. Then, $Q_{ds}$ of each node or community is $Q_{ds}=-SP$.
Therefore, in order to increase $Q_{ds}$ the most, the greedy algorithm would first merge the connected pair of nodes with the sum of their degrees being the largest among all connected pairs. However, it is very likely that those two nodes belong to two different communities, which would finally result in merging those two communities instead of keeping them separate. This will result in a much lower value of $Q_{ds}$ for such a merged community compared to $Q_{ds}$ for its components, demonstrating the reason for greedy $Q_{ds}$ algorithm failure in optimizing $Q_{ds}$.

For example, in the network example in Figure~\ref{two_clique_communities}, the initial values of $Q_{ds}$ for nodes $1, 2, 4, 6, 7,~\text{and}~8$ with degree $3$ are $Q_{ds}=-SP=-\frac{3}{26}$ while the initial values of $Q_{ds}$ for nodes $3$ and $5$ with degree $4$ are $Q_{ds}=-SP=-\frac{4}{26}$. Then, greedy $Q_{ds}$ algorithm would first merge node $3$ and node $5$, which would finally lead to a single community of the whole eight nodes. However, the true community structure contains two clique communities. Accordingly, the $Q_{ds}$ of the community structure with two clique communities, $0.4183$, is larger than that of the community structure with one single large community, $0.2487$. So, maximizing $Q_{ds}$ properly should have the ability to discover the true community structure.

\subsection{Fine-tuned Algorithm}
In this part, we describe a fine-tuned community detection algorithm that iteratively improves a community quality metric $M$ by splitting and merging the given network community structure. We denote the corresponding algorithm based on modularity ($Q$) as \textit{Fine-tuned} $Q$ and the one based on \textit{Modularity Density} ($Q_{ds}$) as \textit{Fine-tuned} $Q_{ds}$. It consists of two alternating stages: split stage and merging stage.

\begin{algorithm}
\caption{Split\_Communities($G$, $C$)}
\label{algorithm:split}
\begin{algorithmic}[1]
    \STATE Initialize comWeights[$|C|$][$|C|$], comEdges[$|C|$][$|C|$], and comDensities[$|C|$][$|C|$] which respectively contain \#weights, \#edges, and the density inside the communities and between two communities by using the network $G$ and the community list $C$;
    \STATE //Get the metric value for each community.
    \STATE $Mes[|C|]$ = GetMetric($C$,comWeights,comDensities);
    \FOR{$i=0$ to $|C|-1$}
      \STATE $c$ = $C$.get($i$);
      \STATE subnet = GenerateSubNetwork($c$);
      \STATE fiedlerVector[$|c|$] = LanczosMethod(subnet);
      \STATE nodeIds[$|c|$] = sort(fiedlerVector, 'descend');
      \STATE //Form $|c|+1$ divisions and record the best one.
      \STATE splitTwoCom.addAll(nodeIds);
      \FOR{$j=0$ to $|c|-1$}
        \STATE splitOneCom.add(nodeIds[$j$]);
        \STATE splitTwoCom.remove(nodeIds[$j$]);
        \STATE Calculate $M(split)$ for the split at $j$;
        \STATE $\Delta M=M(split)-Mes[i]$;
        \IF{$\Delta M(best) < \Delta M$~(or $\Delta M(best) > \Delta M$)}
          \STATE $\Delta M(best) = \Delta M$;
          \STATE $bestIdx = j$;
        \ENDIF
      \ENDFOR

      \IF{$\Delta M(best) > 0$ (or $\Delta M(best)<0$)}
        \STATE Clear splitOneCom and splitTwoCom;
        \STATE splitOneCom.addAll(nodeIds[0:$bestIdx$]);
        \STATE splitTwoCom.addAll(nodeIds[$bestIdx+1$:$|c|-1$);
        \STATE newC.add(splitOneCom);
        \STATE newC.add(splitTwoCom);
      \ELSE
        \STATE newC.add($c$);
      \ENDIF
    \ENDFOR

    \RETURN newC
\end{algorithmic}
\end{algorithm}

In the split stage, the algorithm will split a community $c$ into two subcommunities $c_1$ and $c_2$ based on the ratio-cut method if the split improves the value of the quality metric. The ratio-cut method \cite{RatioCut} finds the bisection that minimizes the ratio $\frac{|E_{c_1,c_2}|}{|c_1||c_2|}$, where $|E_{c_1,c_2}|$ is the cut size (namely, the number of edges between communities $c_1$ and $c_2$), while $|c_1|$ and $|c_2|$ are sizes of the two communities. This ratio penalizes situations in which either of the two communities is small and thus favors balanced divisions over unbalanced ones. However, graph partitioning based on the ratio-cut method is a NP-complete problem. Thus, we approximate it by using the Laplacian spectral bisection method for graph partitioning introduced by Fiedler \cite{Fiedler,FiedlerSparse}.

First, we calculate the \textit{Fiedler vector} which is the eigenvector of the network Laplacian matrix $\bm{L}=D-A$ corresponding to the second smallest eigenvalue. Then, we put the nodes corresponding to the positive values of the \textit{Fiedler vector} into one group and the nodes corresponding to the negative values into the other group. The subnetwork of each community is generated with the nodes and edges in that community. Although the ratio-cut approximated with spectral bisection method does allow some deviation for the sizes $|c_1|$ and $|c_2|$ to vary around the middle value, the right partitioning may not actually divide the community into two balanced or nearly balanced ones. Thus, it is to some extent inappropriate and unrealistic for community detection problems. We overcome this problem by using the following strategies. First, we sort the elements of the \textit{Fiedler vector} in descending order, then cut them into two communities in each of the $|c|+1$ possible ways and calculate the corresponding change of the metric values $\Delta M$ of all the $|c|+1$ divisions. Then, the one with the best value (largest or smallest depending on the measurement) of the quality metric $\Delta M(best)$ among all the $|c|+1$ divisions is recorded. We adopt this best division to the community $c$ only when $\Delta M(best)>0$ (or $\Delta M(best)<0$ depending on the metric). For instance, we split the community only when $\Delta Q_{ds}(best)$ is larger than zero.

The outline of the split stage is shown in Algorithm~\ref{algorithm:split}. The input is a network and a community list, and the output is a list of communities after splitting. The initialization part has $O(|E|)$ complexity. Computing \textit{Fiedler vector} using Lanczos method \cite{Lanczos} needs $O(|E|Kh+|V|K^2h+K^3h)$ steps, where $K$ is the number of eigenvectors needed and $h$ is the number of iterations required for the Lanczos method to converge. Here, $K$ is 2 and $h$ is typically very small although the exact number is not generally known. So, the complexity for calculating \textit{Fiedler vector} is $O(|E|+|V|)$. Sorting the \textit{Fiedler vector} has the cost $O(|V|log|V|)$. The search of the best division from all the $|c|+1$ possible ones (per community $c$) for all the communities is achieved in $O(|E|)$ time. For the $|c|+1$ possible divisions of a community $c$, each one differs from the previous one by the movement of just a single node from one group to the other. Thus, the update of the total weights, the total number of edges, and the densities inside those two split communities and between those two communities to other communities can be calculated in time proportional to the degree of that node. Thus, all nodes can be moved in time proportional to the sum of their degrees which is equal to $2|E|$. Moreover, for \textit{Fine-tuned} $Q_{ds}$, computing $Q_{ds}(split)$ costs $O(|C||V|)$ because all the communities are traversed to update the \textit{Split Penalty} for each of the $|c|+1$ divisions of each community $c$. All the other parts have complexity less than or at most $O(|V|)$. Thus, the computational complexity for the split stage of \textit{Fine-tuned} $Q$ is $O(|E|+|V|log|V|)$ while for \textit{Fine-tuned} $Q_{ds}$ it is $O(|E|+|V|log|V|+|C||V|)$.

\begin{algorithm}
\caption{Merge\_Communities($G$, $C$)}
\label{algorithm:merge}
\begin{algorithmic}[1]
    \STATE Initialize comWeights[$|C|$][$|C|$], comEdges[$|C|$][$|C|$], and comDensities[$|C|$][$|C|$]; 
    \STATE //Get the metric value for each community.
    \STATE $Mes[|C|]$ = GetMetric($C$,comWeights,comDensities);
    \FOR{$i=0$ to $|C|-1$}
      \FOR{$j=i+1$ to $|C|-1$}
        \STATE //Doesn't consider disconnected communities.
        \IF{comWeights[$i$][$j$]==0 \&\& \\
            ~~~~~comWeights[$j$][$i$]==0}
          \STATE continue;
        \ENDIF
        \STATE Calculate $M(merge)$ for merging $c_i$ and $c_j$;
        \STATE $\Delta M = M(merge) - Mes[i] - Mes[j]$;
        \STATE //Record the merging information with $|\Delta M|$ descending in a red-black tree
        \IF{$\Delta M > 0$ (or $\Delta M < 0$)}
          \STATE mergedInfos.put([$|\Delta M|$, $i$, $j$]);
        \ENDIF
      \ENDFOR
    \ENDFOR
    \STATE //Merge the community with the one that improves the value of the quality metric the most
    \WHILE{mergedInfos.hasNext()}
      \STATE [$\Delta M$, comId1, comId2]=mergedInfos.next();
      \IF{!mergedComs.containsKey(comId1) \&\& \\
      ~~~~~!mergedComs.containsKey(comId2)}
        \STATE mergedComs.put(comId1,comId2);
        \STATE mergedComs.put(comId2,comId1);
      \ENDIF
    \ENDWHILE
    \FOR{$i=0$ to $|C|-1$}
      \STATE $c_i$=$C$.get($i$);
      \IF{mergedComs.containsKey($i$)}
        \STATE comId2 = mergedComs.get($i$);
        \IF{$i < $ comId2}
          \STATE $c_i$.addAll($C$.get(comId2));
        \ENDIF
      \ENDIF
      \STATE newC.add($c_i$);
    \ENDFOR
    \RETURN newC;
\end{algorithmic}
\end{algorithm}

In the merging stage, the algorithm will merge a community to its connected communities if the merging improves the value of the quality metric. If there are many mergers possible for a community, the one, unmerged so far, which improves the quality metric the most is chosen. Hence, each community will only be merged at most once in each stage. The outline of the merging stage is shown Algorithm~\ref{algorithm:merge}. The input is a network and a community list, and the output is a list of communities after merging. The initialization part has the complexity $O(|E|)$. For \textit{Fine-tuned} $Q$, the two ``for loops'' for merging any two communities have the complexity $O(|C|^2log|C|)$ because calculating $Q(merge)$ is $O(1)$ and inserting an element into the red-black tree is $O(log|C|^2)=O(2log|C|) \sim O(log|C|)$ since the maximum number of elements in the tree is $\frac{|C|(|C|-1)}{2}=O(|C|^2)$. For \textit{Fine-tuned} $Q_{ds}$, the two ``for loops'' for merging any two communities have the complexity $O(|C|^3)$ because calculating $Q_{ds}(merge)$ needs $O(|C|)$ steps to traverse all the communities to update the \textit{Split Penalty} and inserting an element into the red-black tree is $O(log|C|)$ as well. The other parts all have complexity at most $O(|C|^2)$. Thus, the computational complexity for the merging stage of \textit{Fine-tuned} $Q$ is $O(|E|+|C|^2log|C|)$ and for the merging stage of \textit{Fine-tuned} $Q_{ds}$ is $O(|E|+|C|^3)$.

\begin{algorithm}
\caption{Fine-tuned\_Algorithm($G$, $C$)}
\label{algorithm:fine-tuned}
\begin{algorithmic}[1]
    \STATE comSize = $|C|$;
    \STATE splitSize = 0;
    \STATE mergeSize = 0;
    \WHILE{comSize!=splitSize $\|$ comSize!=mergeSize}
      \STATE comSize = $|C|$;
      \STATE $C$ = Split\_Communities($G$, $C$);
      \STATE splitSize = $|C|$;
      \STATE $C$=Merge\_Communities($G$, $C$);
      \STATE mergeSize = $|C|$;
    \ENDWHILE
    \RETURN $C$
\end{algorithmic}
\end{algorithm}

The fine-tuned algorithm repeatedly carries out those two alternating stages until neither split nor merging can improve the value of the quality metric or until the total number of communities discovered does not change after one full iteration. Algorithm~\ref{algorithm:fine-tuned} shows the outline of the fine-tuned algorithm. It can detect the community structure of a network by taking a list with a single community of all the nodes in the network as the input. It can also improve the community detection results of other algorithms by taking a list with their communities as the input. Let the number of iteration of the fine-tuned algorithm be denoted as $T$. Then, the total complexity for \textit{Fine-tuned} $Q$ is $O(T(|E|+|V|log|V|+|C|^2log|C|))$ while for \textit{Fine-tuned} $Q_{ds}$ it is $O(T(|E|+|V|log|V|+|C||V|+|C|^3))$. Assuming that $T$ and $|C|$ are constants, the complexity of the fine-tuned algorithms reduces to $O(|E|+|V|log|V|)$. The only part of the algorithm that would generate a non-deterministic result is the Lanczos method of calculating the \textit{Fiedler vector}. The reason is that Lanczos method adopts a randomly generated vector as its starting vector. We solve this issue by choosing a normalized vector of the size equal to the number of nodes in the community as the starting vector for the Lanczos method. Then, community detection results will stay the same for different runs as long as the input remains the same.

\section{Experimental Results}
In this section, we first introduce several popular measurements for evaluating the quality of the results of community detection algorithms. Denoting the greedy algorithm of modularity maximization proposed by Newman \cite{ModularityLargeNet} as \textit{Greedy} $Q$, we then use the mentioned above metrics to compare \textit{Greedy} $Q$, \textit{Fine-tuned} $Q$, and \textit{Fine-tuned} $Q_{ds}$. The comparison uses four real networks, the classical clique network and the LFR benchmark networks, each instance of which is defined with parameters each selected from a wide range of possible values. The results indicate that \textit{Fine-tuned} $Q_{ds}$ is the most effective method among the three, followed by \textit{Fine-tuned} $Q$. Moreover, we show that \textit{Fine-tuned} $Q_{ds}$ can be applied to significantly improve the detection results of other algorithms.

In Subsection~\ref{spectral_method}, we have shown that the modularity maximization approach using the eigenvectors of the Laplacian matrix is equivalent to the one using the eigenvectors of the modularity matrix. This implies that the split stage of our \textit{Fine-tuned} $Q$ is actually equivalent to the spectral methods. Therefore, \textit{Fine-tuned} $Q$ with one additional merge operation at each iteration unquestionably has better performance than the spectral algorithms. Hence, we do not discuss them here.

\subsection{Evaluation Metrics}
\label{evaluation_metrics}
The quality evaluation metrics we consider here can be divided into three categories: \textit{Variation of Information} ($VI$) \cite{ExtremeDegeneracy} and \textit{Normalized Mutual Information} ($NMI$) \cite{NMI} based on information theory; \textit{F-measure} \cite{F_measure} and \textit{Normalized Van Dongen metric} ($NVD$) \cite{NVD} based on cluster matching; \textit{Rand Index} ($RI$) \cite{RandIndex}, \textit{Adjusted Rand Index} ($ARI$) \cite{AdjustedRandIndex}, and \textit{Jaccard Index} ($JI$) \cite{JaccardIndex} based on pair counting.

\begin{table*}[!t]
\caption{Metric values of the community structures discovered by \textit{Greedy} $Q$, \textit{Fine-tuned} $Q$, and \textit{Fine-tuned} $Q_{ds}$ on Zachary's karate club network (red italic font denotes the best value for each metric).}
\label{karate_table}
\vspace{-1.2em}
\centering
\setlength{\tabcolsep}{10pt}
\begin{tabular}{c||c|c|c|c|c|c|c|c|c}
\hline \hline
     ~~~Algorithm~~~ & $Q$ & $Q_{ds}$ & $VI$ & $NMI$ & $F\textit{-}measure$ &	$NVD$ & $RI$ &	$ARI$ & $JI$ \\
\hline
     Greedy $Q$ & 0.3807 & 0.1809 &	\textcolor{red}{\textbf{\emph{0.7677}}} & 0.6925 & \textcolor{red}{\textbf{\emph{0.828}}} & \textcolor{red}{\textbf{\emph{0.1471}}} & \textcolor{red}{\textbf{\emph{0.8414}}} & \textcolor{red}{\textbf{\emph{0.6803}}} & \textcolor{red}{\textbf{\emph{0.6833}}} \\
\hline
     Fine-tuned $Q$ &  \textcolor{red}{\textbf{\emph{0.4198}}} & 0.2302 & 0.9078 & 0.6873 &	0.807 & 0.1618 & 0.7736 & 0.5414 &	0.5348\\
\hline
     Fine-tuned $Q_{ds}$ &  0.4174 & \textcolor{red}{\textbf{\emph{0.231}}} & 0.8729 & \textcolor{red}{\textbf{\emph{0.6956}}} & 0.8275 & \textcolor{red}{\textbf{\emph{0.1471}}} & 0.7861 & 0.5669 & 0.5604\\
\hline \hline
\end{tabular}
\vspace{1em}
\end{table*}

\begin{figure*}[!t]
\centering
\subfigure[Ground truth communities.]{
\label{karate:subfig:a}
\includegraphics[scale=0.59, clip]{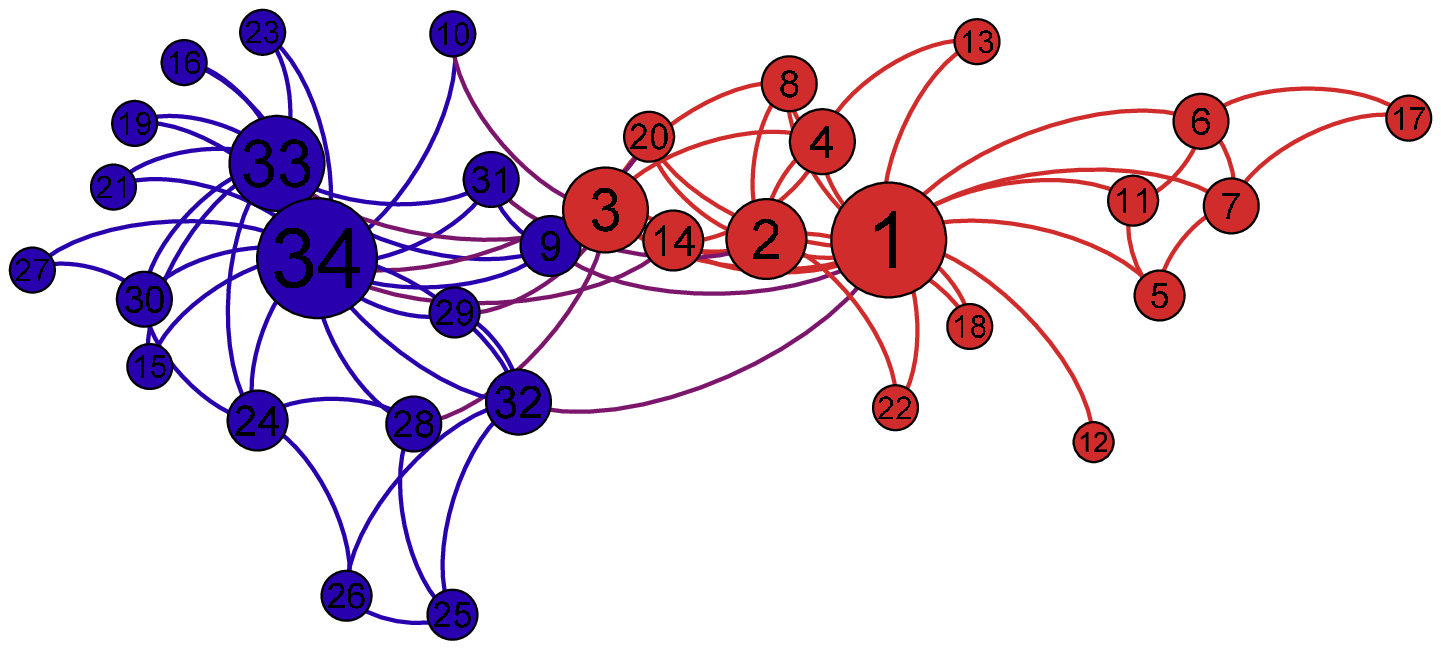}
}
\hspace{1em}
\subfigure[Communities detected with \textit{Greedy} $Q$.]{
\label{karate:subfig:b}
\includegraphics[scale=0.42, clip]{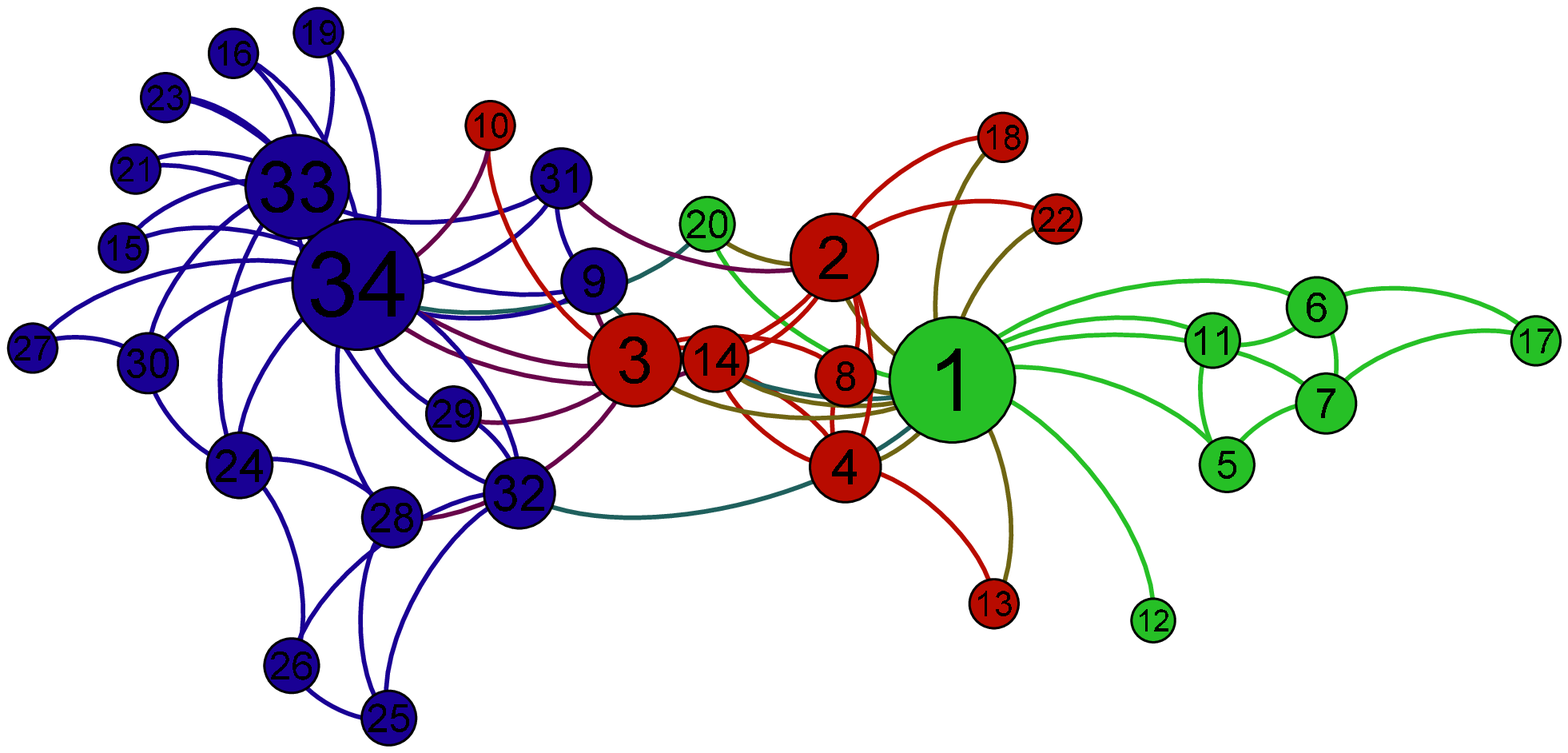}
}
\vspace{1em}
\subfigure[Communities detected with \textit{Fine-tuned} $Q$.]{
\label{karate:subfig:c}
\includegraphics[scale=0.44, clip]{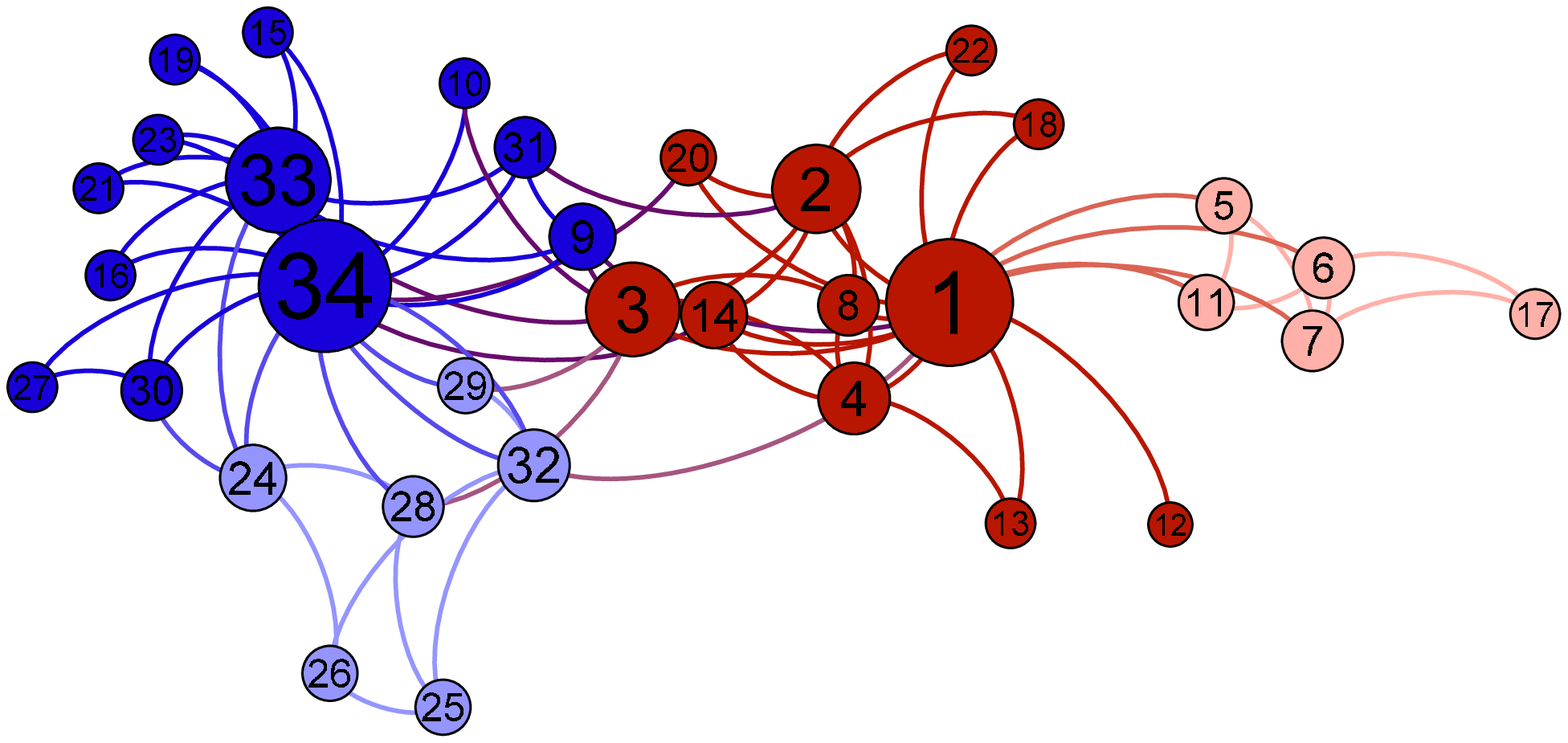}
}
\hspace{1em}
\subfigure[Communities detected with \textit{Fine-tuned} $Q_{ds}$.]{
\label{karate:subfig:d}
\includegraphics[scale=0.44, clip]{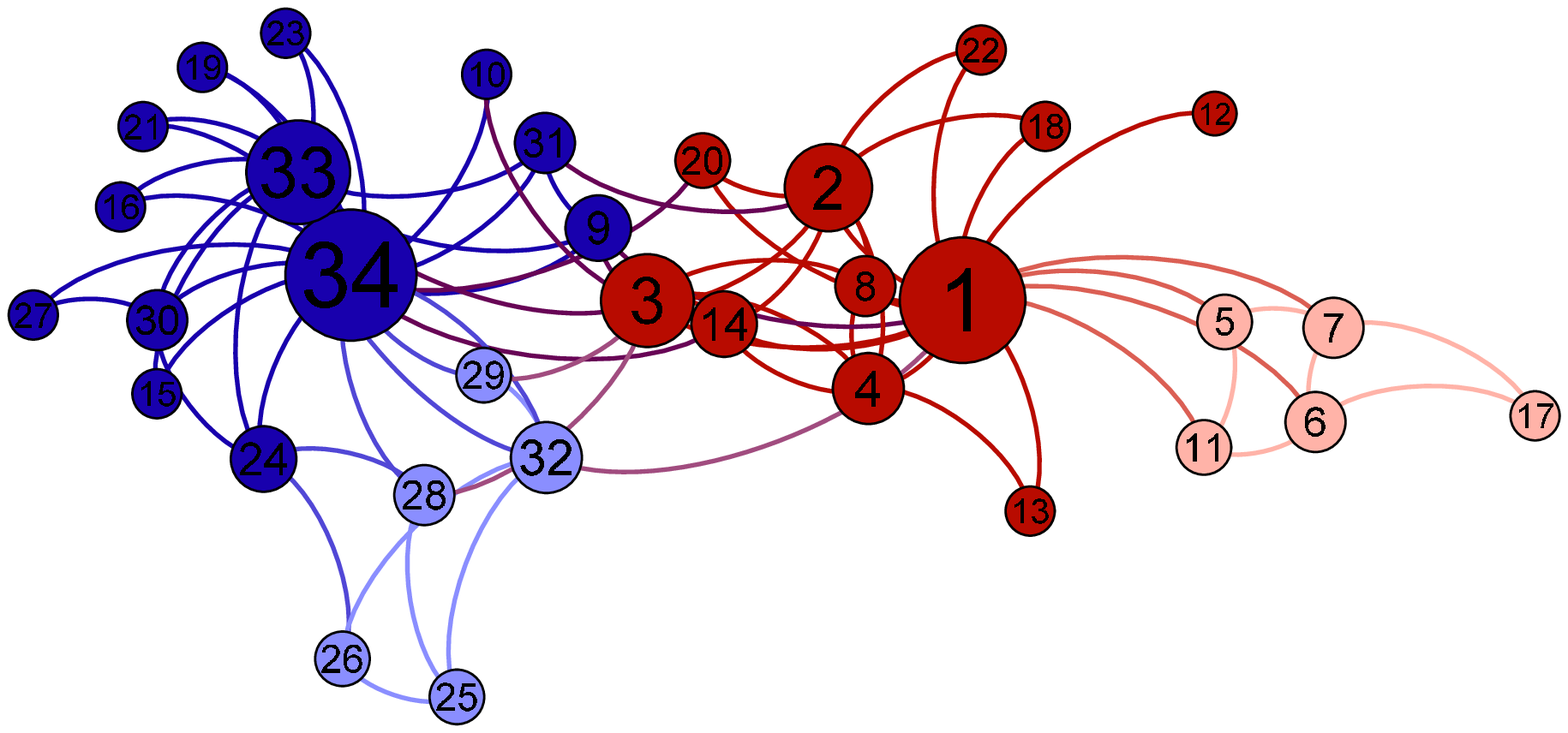}
}
\vspace{-1.9em}
\centering
\caption{The community structures of the ground truth communities and those detected by \textit{Greedy} $Q$, \textit{Fine-tuned} $Q$, and \textit{Fine-tuned} $Q_{ds}$ on Zachary's karate club network.}
\label{karate_figure}
\vspace{-0.5em}
\end{figure*}

\subsubsection{Information Theory Based Metrics}
Given partitions $C$ and $C'$, \textit{Variation of Information} ($VI$) \cite{ExtremeDegeneracy} quantifies the ``distance'' between those two partitions, while \textit{Normalized Mutual Information} ($NMI$) \cite{NMI} measures the similarity between partitions $C$ and $C'$. $VI$ is defined as
\begin{equation}
\begin{split}
VI(C,C')&=H(C)+H(C')-2I(C,C') \\
&=H(C,C')-I(C,C'),
\end{split}
\end{equation}
where $H(.)$ is the entropy function and $I(C,C')=H(C)+H(C')-H(C,C')$ is the \textit{Mutual Information}. Then, $NMI$ is given by
\begin{equation}
NMI(C,C')=\frac{2I(C,C')}{H(C)+H(C')}.
\end{equation}
Using the definitions
\begin{equation}
H(C)=-\sum_{c_i \in C}p(c_i)\log p(c_i)=-\sum_{c_i \in C}\frac{|c_i|}{|V|}\log \frac{|c_i|}{|V|},
\end{equation}
\begin{equation}
\begin{split}
H(C,C')&=-\sum_{c_i \in C,c_j' \in C'}p(c_i,c_j')\log p(c_i,c_j')  \\
&=-\sum_{c_i \in C,c_j' \in C'}\frac{|c_i \cap c_j'|}{|V|}\log \left(\frac{|c_i \cap c_j'|}{|V|}\right)
\end{split}
\end{equation}
we can express $VI$ and $NMI$ as a function of counts only as follows:
\begin{equation}
VI(C,C')=-\frac{1}{|V|}\sum_{c_i \in C, c_j' \in C'} |c_i \cap c_j'| \log \left(\frac{|c_i \cap c_j'|^2}{|c_i||c_j'|}\right),
\end{equation}
\begin{equation}
NMI(C,C')=\frac{-2\sum_{c_i \in C, c_j' \in C'} \frac{|c_i \cap c_j'|}{|V|} \log \left(\frac{|c_i \cap c_j'||V|}{|c_i||c_j'|}\right)}{\sum_{c_i \in C}\frac{|c_i|}{|V|}\log \frac{|c_i|}{|V|}+\sum_{c_j' \in C'}\frac{|c_j'|}{|V|}\log \frac{|c_j'|}{|V|}},
\end{equation}
where $|c_i|$ is the number of nodes in community $c_i$ of $C$ and $|c_i \cap c_j'|$ is the number of nodes both in community $c_i$ of $C$ and in community $c_j'$ of $C'$.

\subsubsection{Clustering Matching Based Metrics}
Measurements based on clustering matching aim at finding the largest overlaps between pairs of communities of two partitions $C$ and $C'$. \textit{F-measure} \cite{F_measure} measures the similarity between two partitions, while \textit{Normalized Van Dongen metric} ($NVD$) \cite{NVD} quantifies the ``distance'' between partitions $C$ and $C'$. \textit{F-measure} is defined as
\begin{equation}
F\textit{-}measure(C,C')=\frac{1}{|V|}\sum_{c_i \in C} |c_i| \max_{c_j' \in C'} \frac{2|c_i \cap c_j'|}{|c_i|+|c_j'|}.
\end{equation}
$NVD$ is given by
\begin{equation}
\begin{split}
&NVD(C,C')=1-\frac{1}{2|V|} \biggl(\sum_{c_i \in C} \max_{c_j' \in C'} |c_i \cap c_j'|  \\
&~~~~~~~~~~~~~~~~~~~+\sum_{c_j' \in C'} \max_{c_i \in C} |c_j' \cap c_i| \biggr).
\end{split}
\end{equation}

\subsubsection{Pair Counting Based Metrics}
Metrics based on pair counting count the number of pairs of nodes that are classified (in the same community or in different communities) in two partitions $C$ and $C'$. Let $a_{11}$ indicate the number of pairs of nodes that are in the same community in both partitions, $a_{10}$ denote the number of pairs of nodes that are in the same community in partition $C$ but in different communities in $C'$, $a_{01}$ be the number of pairs of nodes which are in different communities in $C$ but in the same community in $C'$, $a_{00}$ be the number of pairs of nodes which are in different communities in both partitions. By definition, $A=a_{11}+a_{10}+a_{01}+a_{00}=\frac{|V|(|V|-1)}{2}$ is the total number of pairs of nodes in the network. Then, \textit{Rand Index} ($RI$) \cite{RandIndex} which is the ratio of the number of node pairs placed in the same way in both partitions to the total number of pairs is given by
\begin{equation}
RI(C,C')=\frac{a_{11}+a_{00}}{A}.
\end{equation}
Denote $M=\frac{1}{A}(a_{11}+a_{10})(a_{11}+a_{01})$. Then, \textit{RI}'s corresponding adjusted version, \textit{Adjusted Rand Index} ($ARI$) \cite{AdjustedRandIndex}, is expressed as
\begin{equation}
ARI(C,C')=\frac{a_{11}-M}{\frac{1}{2}\left[(a_{11}+a_{10})+(a_{11}+a_{01})\right]-M}.
\end{equation}
The \textit{Jaccard Index} ($JI$) \cite{JaccardIndex} which is the ratio of the number of node pairs placed in the same community in both partitions to the number of node pairs that are placed in the same group in at least one partition is defined as
\begin{equation}
JI(C,C')=\frac{a_{11}}{a_{11}+a_{10}+a_{01}}.
\end{equation}
Each of these three metrics quantifies the similarity between two partitions $C$ and $C'$.

\begin{table*}[!t]
\caption{Metric values of the community structures detected by \textit{Greedy} $Q$, \textit{Fine-tuned} $Q$, and \textit{Fine-tuned} $Q_{ds}$ on American college football network (red italic font denotes the best value for each metric).}
\label{football_table}
\vspace{-1.2em}
\centering
\setlength{\tabcolsep}{10pt}
\begin{tabular}{c||c|c|c|c|c|c|c|c|c}
\hline \hline
     ~~~Algorithm~~~ & $Q$ & $Q_{ds}$ & $VI$ & $NMI$ & $F\textit{-}measure$ &	$NVD$ & $RI$ &	$ARI$ & $JI$ \\
\hline
     Greedy $Q$ & 0.5773 & 0.3225 &	1.4797 & 0.7624 & 0.6759 & 0.2304 &	0.9005 & 0.5364 & 0.4142 \\
\hline
     Fine-tuned $Q$ &  0.5944 & 0.3986 & 0.9615 & 0.8553 & 0.8067 & 0.1348 & 0.9521 & 0.7279 & 0.6045\\
\hline
     Fine-tuned $Q_{ds}$ &  \textcolor{red}{\textbf{\emph{0.6005}}} & \textcolor{red}{\textbf{\emph{0.4909}}} & \textcolor{red}{\textbf{\emph{0.5367}}} & \textcolor{red}{\textbf{\emph{0.9242}}} & \textcolor{red}{\textbf{\emph{0.9145}}} & \textcolor{red}{\textbf{\emph{0.07391}}} & \textcolor{red}{\textbf{\emph{0.9847}}} & \textcolor{red}{\textbf{\emph{0.8967}}} & \textcolor{red}{\textbf{\emph{0.8264}}} \\
\hline \hline
\end{tabular}
\vspace{-0.2em}
\end{table*}

\begin{table*}[!t]
\caption{Metric values of the community structures of \textit{Greedy} $Q$ and \textit{Fine-tuned} $Q$ improved with \textit{Fine-tuned} $Q_{ds}$ on American college football network (blue italic font indicates improved score).}
\label{football_improved}
\vspace{-1.2em}
\centering
\setlength{\tabcolsep}{6pt}
\begin{tabular}{c||c|c|c|c|c|c|c|c|c}
\hline \hline
     ~~Algorithm~~ & $Q$ & $Q_{ds}$ & $VI$ & $NMI$ & $F\textit{-}measure$ &	$NVD$ & $RI$ &	$ARI$ & $JI$ \\
\hline
     Greedy $Q$ improved with Fine-tuned $Q_{ds}$ & \textcolor{blue}{\textbf{\emph{0.5839}}} & \textcolor{blue}{\textbf{\emph{0.4636}}} & \textcolor{blue}{\textbf{\emph{0.6986}}} & \textcolor{blue}{\textbf{\emph{0.9013}}} & \textcolor{blue}{\textbf{\emph{0.8961}}} & \textcolor{blue}{\textbf{\emph{0.0913}}} & \textcolor{blue}{\textbf{\emph{0.9793}}} & \textcolor{blue}{\textbf{\emph{0.8597}}} & \textcolor{blue}{\textbf{\emph{0.7714}}} \\
\hline
     Fine-tuned $Q$ improved with Fine-tuned $Q_{ds}$ & \textcolor{blue}{\textbf{\emph{0.5974}}} & \textcolor{blue}{\textbf{\emph{0.4793}}} & \textcolor{blue}{\textbf{\emph{0.5096}}} & \textcolor{blue}{\textbf{\emph{0.9278}}} & \textcolor{blue}{\textbf{\emph{0.9166}}} & \textcolor{blue}{\textbf{\emph{0.06957}}} & \textcolor{blue}{\textbf{\emph{0.9837}}} & \textcolor{blue}{\textbf{\emph{0.8907}}} & \textcolor{blue}{\textbf{\emph{0.8174}}} \\
\hline \hline
\end{tabular}
\vspace{-0.2em}
\end{table*}

\begin{figure*}[!t]
\centering
\setlength{\belowcaptionskip}{-1em}
\subfigure[Ground truth communities.]{
\label{football:subfig:a}
\includegraphics[scale=0.42, clip]{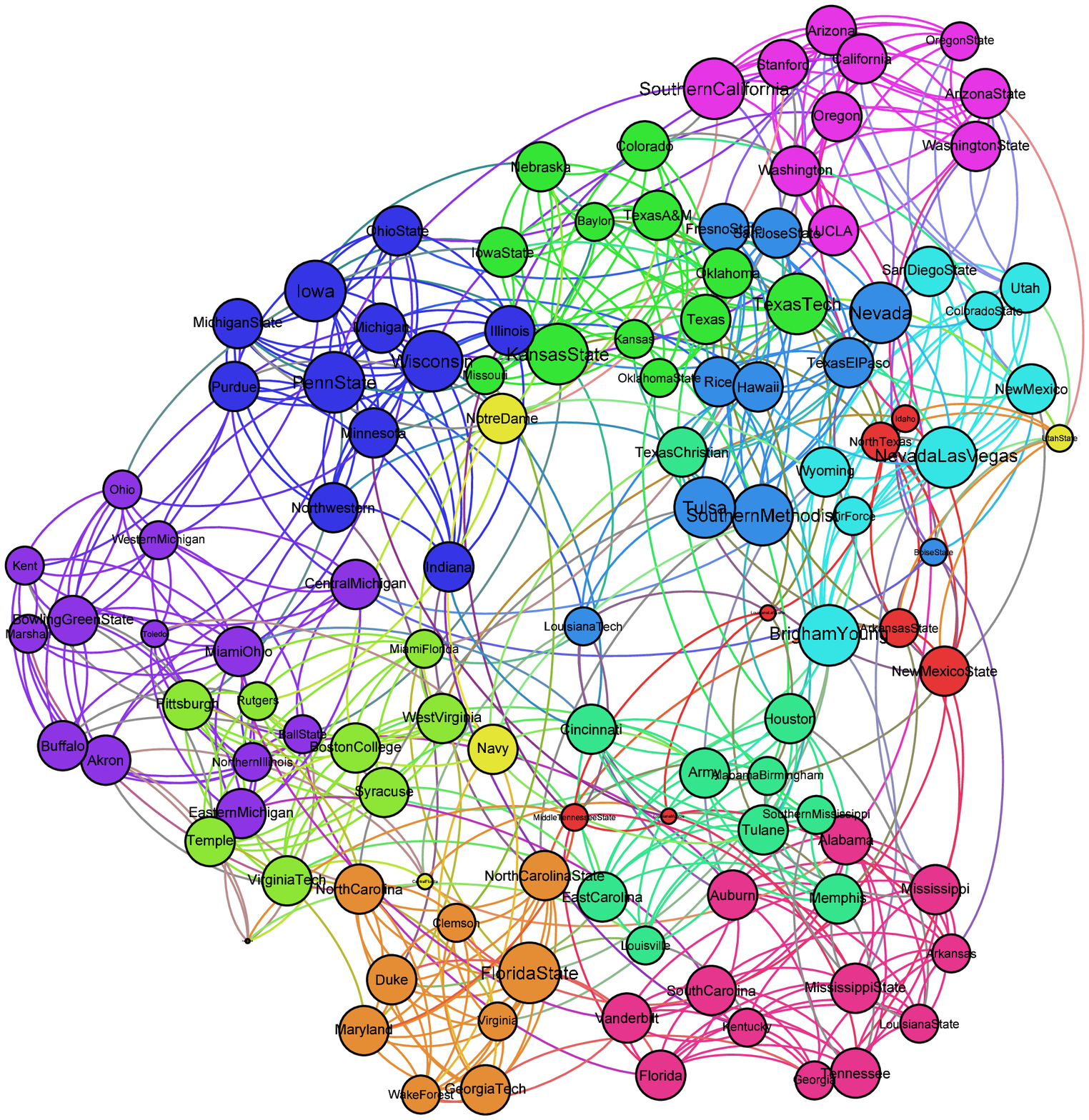}
}
\hspace{0.5em}
\subfigure[Communities detected with \textit{Greedy} $Q$.]{
\label{football:subfig:b}
\includegraphics[scale=0.41, clip]{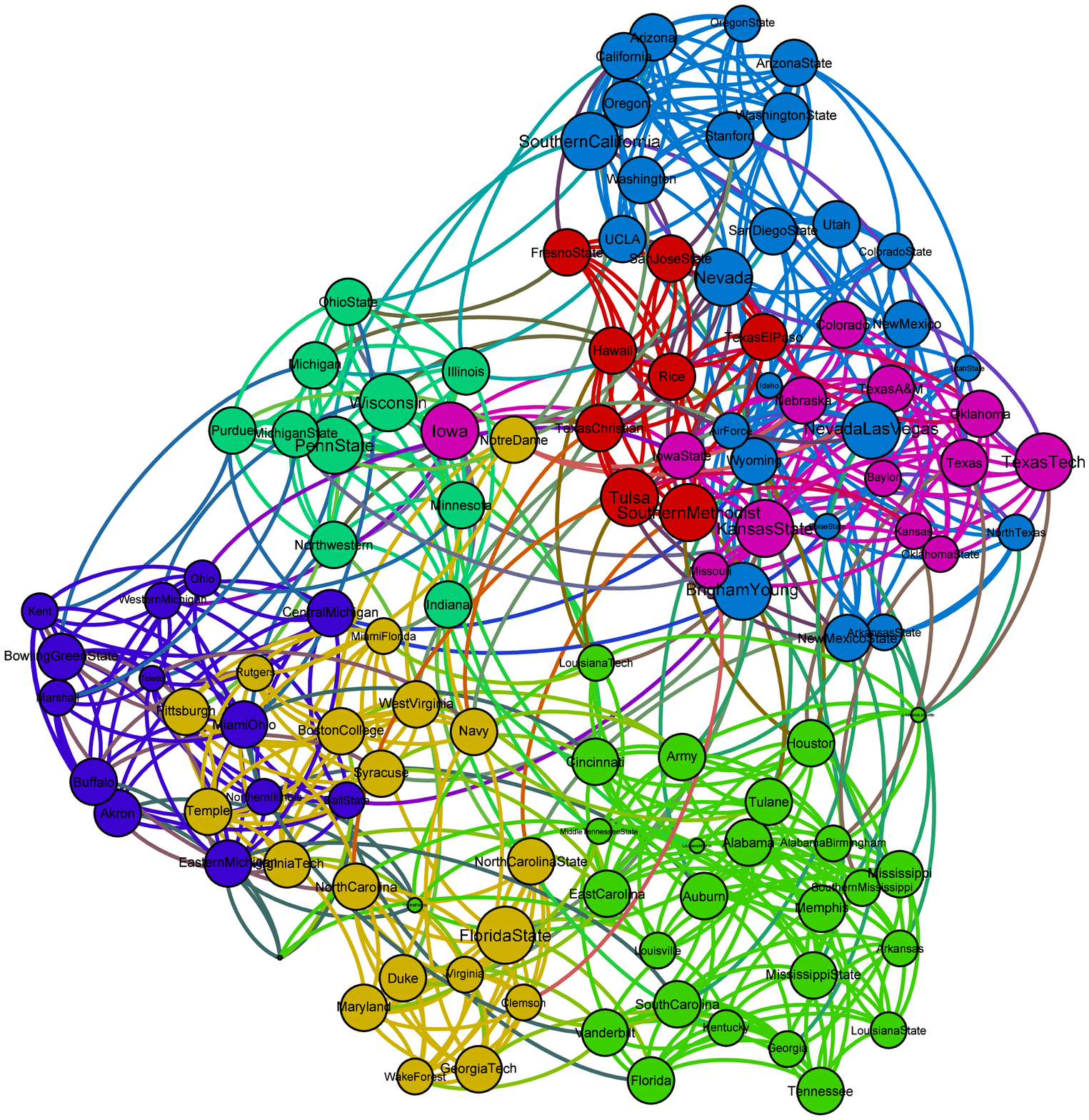}
}
\vspace{1em}
\subfigure[Communities detected with \textit{Fine-tuned} $Q$.]{
\label{football:subfig:c}
\includegraphics[scale=0.41, clip]{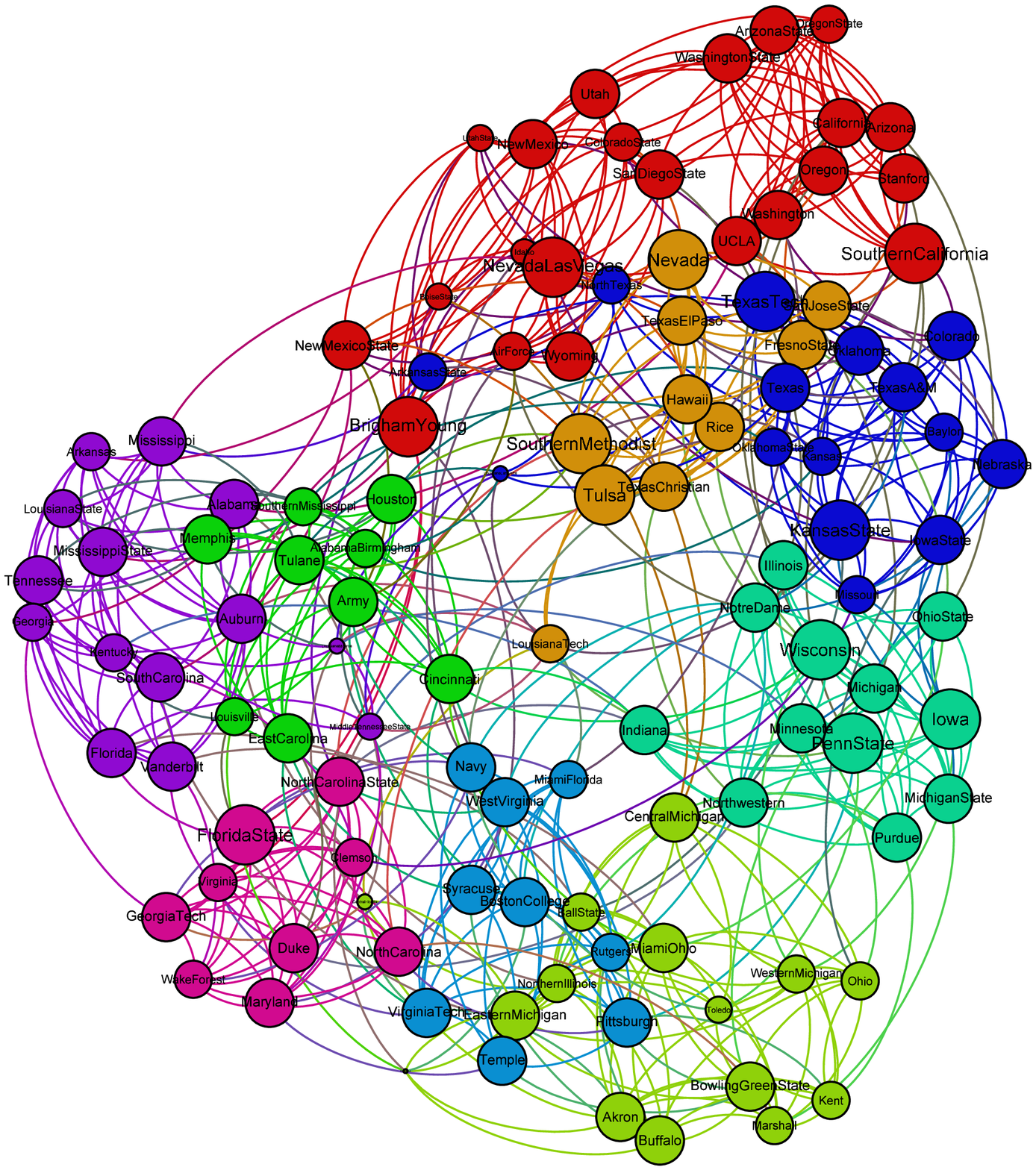}
}
\hspace{0.5em}
\subfigure[Communities detected with \textit{Fine-tuned} $Q_{ds}$.]{
\label{football:subfig:d}
\includegraphics[scale=0.44, clip]{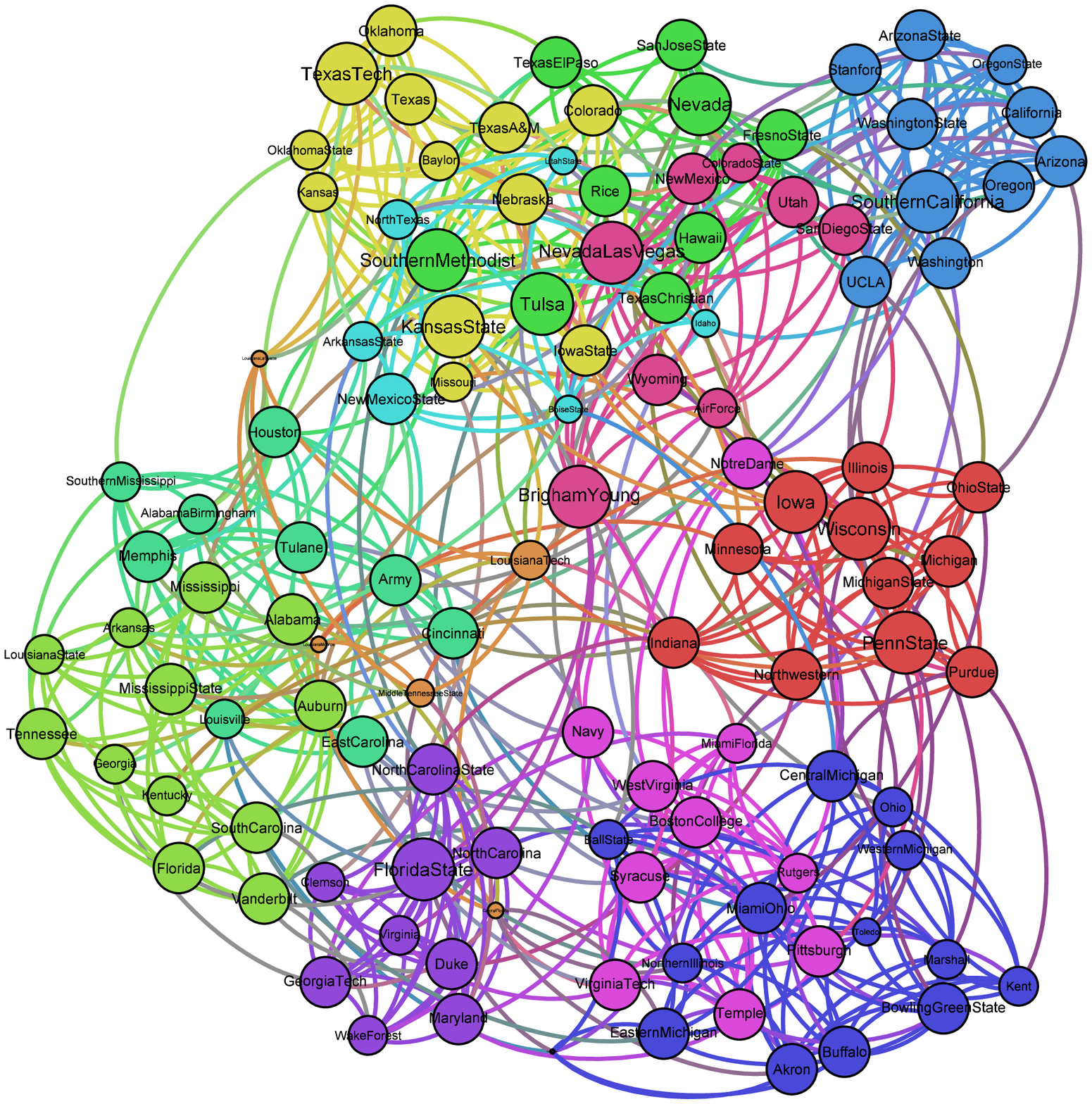}
}
\vspace{-1.8em}
\centering
\caption{The community structures of the ground truth communities and those detected by \textit{Greedy} $Q$, \textit{Fine-tuned} $Q$, and \textit{Fine-tuned} $Q_{ds}$ on American college football network.}
\label{football_figure}
\vspace{-0.5em}
\end{figure*}

\subsection{Real Networks}
In this subsection, we first evaluate the performance of \textit{Greedy} $Q$, \textit{Fine-tuned} $Q$, and \textit{Fine-tuned} $Q_{ds}$ on two small networks (Zachary's karate club network \cite{karate} and American college football network \cite{football}) with ground truth communities, and then on two large networks (PGP network \cite{PGPNetwork} and AS level Internet) but without ground truth communities.

\subsubsection{Zachary's Karate Club Network}
We first compare the performance of \textit{Greedy} $Q$, \textit{Fine-tuned} $Q$, and \textit{Fine-tuned} $Q_{ds}$ on Zachary's karate club network \cite{karate}. It represents the friendships between $34$ members of a karate club at a US university over a period of $2$ years. During the observation period, the club split into two clubs as a result of a conflict within the organization. The resulting two new clubs can be treated as the ground truth communities whose structure is shown in Figure~\ref{karate:subfig:a} visualized with the opensource software Gephi \cite{Gephi}.

Table~\ref{karate_table} presents the metric values of the community structures detected by the three algorithms on this network. It shows that \textit{Fine-tuned} $Q$ and \textit{Fine-tuned} $Q_{ds}$ achieve the highest value of $Q$ and $Q_{ds}$, respectively. However, most of the seven metrics based on ground truth communities imply that \textit{Greedy} $Q$ performs the best with only $NMI$ and $NVD$ indicating that \textit{Fine-tuned} $Q_{ds}$ has the best performance among the three algorithms. Hence, it seems that a large $Q$ or $Q_{ds}$ may not necessary mean a high quality of community structure, especially for $Q$ because \textit{Fine-tuned} $Q$ achieves the highest $Q$ but has the worst values of the seven metrics described in Subsection~\ref{evaluation_metrics}. We argue that the ground truth communities may not be so reasonable because \textit{Fine-tuned} $Q$ and \textit{Fine-tuned} $Q_{ds}$ in fact discover more meaningful communities than \textit{Greedy} $Q$ does. Figures~\ref{karate:subfig:a}-\ref{karate:subfig:d} show the community structure of ground truth communities and those detected by \textit{Greedy} $Q$, \textit{Fine-tuned} $Q$, and \textit{Fine-tuned} $Q_{ds}$, respectively. For results of \textit{Greedy} $Q$ shown in Figure~\ref{karate:subfig:b}, we could observe that there are three communities located at the left, the center, and the right side of the network. The ground truth community located on the right is subdivided into the central and right communities, but the node 10 is misclassified as belonging to the central community, while in ground truth network it belongs to community located on the left. Figure~\ref{karate:subfig:c} demonstrates that \textit{Fine-tuned} $Q$ subdivides both the left and the right communities into two with six nodes separated from the left community and five nodes separated from the right community. Moreover, Figure~\ref{karate:subfig:c} shows that \textit{Fine-tuned} $Q$ discovers the same number of communities for this network as algorithms presented in \cite{Louvain,QMaxWithSAMedus,MathProgrammingQ,ExtremalQ}. In fact, the community structure it discovers is identical to those detected in \cite{QMaxWithSAMedus,MathProgrammingQ,ExtremalQ}. Figure~\ref{karate:subfig:d} shows that the community structure discovered by \textit{Fine-tuned} $Q_{ds}$ differs from that of \textit{Fine-tuned} $Q$ only on node 24 which is placed in the larger part of the left community. It is reasonable for it has three connections to the larger part to which it has more attraction than to the smaller part with which it only has two connections.

In addition, analyzing the intermediate results of \textit{Fine-tuned} $Q$ and \textit{Fine-tuned} $Q_{ds}$ reveals that the communities at the first iteration are exactly the ground truth communities, which in another way implies their superiority over \textit{Greedy} $Q$. Moreover, $NMI$ and $NVD$ indicate that \textit{Fine-tuned} $Q_{ds}$ is the best among the three and all the metrics, except $Q$, show that \textit{Fine-tuned} $Q_{ds}$ performs better than \textit{Fine-tuned} $Q$, supporting the claim that a higher $Q_{ds}$ (but not $Q$) implies a better quality of community structure.

\subsubsection{American College Football Network}
We apply the three algorithms also to the American college football network \cite{football} which represents the schedule of games between college football teams in a single season. The teams are divided into twelve ``conferences'' with intra-conference games being more frequent than inter-conference games. Those conferences could be treated as the ground truth communities whose structure is shown in Figure~\ref{football:subfig:a}.

Table~\ref{football_table} presents the metric values of the community structures detected by the three algorithms. It shows that \textit{Fine-tuned} $Q_{ds}$ achieves the best values for all the nine metrics. It implies that \textit{Fine-tuned} $Q_{ds}$ performs best on this football network, followed by \textit{Fine-tuned} $Q$. Figures~\ref{football:subfig:a}-\ref{football:subfig:d} present the community structure of ground truth communities and those discovered by \textit{Greedy} $Q$, \textit{Fine-tuned} $Q$, and \textit{Fine-tuned} $Q_{ds}$. Each color in the figures represents a community. It can be seen that there are twelve ground truth communities in total, seven communities detected by \textit{Greedy} $Q$, nine communities discovered by \textit{Fine-tuned} $Q$, and exactly twelve communities found by \textit{Fine-tuned} $Q_{ds}$.

Moreover, we apply \textit{Fine-tuned} $Q_{ds}$ on the community detection results of \textit{Greedy} $Q$ and \textit{Fine-tuned} $Q$. The metric values of these two community structures after improvement with \textit{Fine-tuned} $Q_{ds}$ are shown in Table~\ref{football_improved}. Compared with those of \textit{Greedy} $Q$ and \textit{Fine-tuned} $Q$ in Table~\ref{football_table}, we could observe that the metric values are significantly improved with \textit{Fine-tuned} $Q_{ds}$. Further, both improved community structures contain exactly twelve communities, the same number as the ground truth communities.

\begin{table}[!t]
\caption{The values of $Q$ and $Q_{ds}$ of the community structures detected by \textit{Greedy} $Q$, \textit{Fine-tuned} $Q$, and \textit{Fine-tuned} $Q_{ds}$ on PGP network (red italic font denotes the best value for each metric).}
\label{pgp_table}
\vspace{-1.2em}
\centering
\setlength{\tabcolsep}{18pt}
\begin{tabular}{c||c|c}
\hline \hline
     ~~Algorithm~~ & $Q$ & $Q_{ds}$ \\
\hline
     Greedy $Q$ & \textcolor{red}{\textbf{\emph{0.8521}}} & 0.04492  \\
\hline
     Fine-tuned $Q$ & 0.8405 & 0.02206  \\
\hline
     Fine-tuned $Q_{ds}$ & 0.594 & \textcolor{red}{\textbf{\emph{0.287}}} \\
\hline \hline
\end{tabular}
\vspace{-0.5em}
\end{table}

\subsubsection{PGP Network}
We then apply the three algorithms on PGP network \cite{PGPNetwork}. It is the giant component of the network of users of the Pretty-Good-Privacy algorithm for secure information interchange. It has 10680 nodes and 24316 edges.

Table~\ref{pgp_table} presents the metric values of the community structures detected by the three algorithms. Since this network does not have ground truth communities, we only calculate $Q$ and $Q_{ds}$ of these discovered community structures. The table shows that \textit{Greedy} $Q$ and \textit{Fine-tuned} $Q_{ds}$ achieve the highest value of $Q$ and $Q_{ds}$, respectively. It is worth to mention that the $Q_{ds}$ of \textit{Fine-tuned} $Q_{ds}$ is much larger than that of \textit{Greedy} $Q$ and \textit{Fine-tuned} $Q$, which implies that \textit{Fine-tuned} $Q_{ds}$ performs best on PGP network according to $Q_{ds}$, followed by \textit{Greedy} $Q$.

\begin{table}[!t]
\caption{The values of $Q$ and $Q_{ds}$ of the community structures detected by \textit{Greedy} $Q$, \textit{Fine-tuned} $Q$, and \textit{Fine-tuned} $Q_{ds}$ on AS level Internet (red italic font denotes the best value for each metric).}
\label{as_table}
\vspace{-1.2em}
\centering
\setlength{\tabcolsep}{18pt}
\begin{tabular}{c||c|c}
\hline \hline
     ~~Algorithm~~ & $Q$ & $Q_{ds}$ \\
\hline
     Greedy $Q$ & 0.6379 & 0.002946  \\
\hline
     Fine-tuned $Q$ & \textcolor{red}{\textbf{\emph{0.6475}}} & 0.003123  \\
\hline
     Fine-tuned $Q_{ds}$ & 0.3437 & \textcolor{red}{\textbf{\emph{0.03857}}} \\
\hline \hline
\end{tabular}
\vspace{-0.2em}
\end{table}

\begin{table*}[!t]
\caption{Metric values of the community structures detected by \textit{Greedy} $Q$, \textit{Fine-tuned} $Q$, and \textit{Fine-tuned} $Q_{ds}$ on the classical clique network (red italic font denotes the best value for each metric).}
\label{clique_table}
\vspace{-1.2em}
\centering
\setlength{\tabcolsep}{10pt}
\begin{tabular}{c||c|c|c|c|c|c|c|c|c}
\hline \hline
     ~~~Algorithm~~~ & $Q$ & $Q_{ds}$ & $VI$ & $NMI$ & $F\textit{-}measure$ &	$NVD$ & $RI$ &	$ARI$ & $JI$ \\
\hline
     Greedy $Q$ & \textcolor{red}{\textbf{\emph{0.8871}}} & 0.46 & 0.9333 & 0.8949 & 0.6889 & 0.2333 & 0.9687 & 0.6175 & 0.4615 \\
\hline
     Fine-tuned $Q$ & \textcolor{red}{\textbf{\emph{0.8871}}} & 0.46 & 0.9333 & 0.8949 & 0.6889 & 0.2333 & 0.9687 & 0.6175 & 0.4615 \\
\hline
     Fine-tuned $Q_{ds}$ & 0.8758 & \textcolor{red}{\textbf{\emph{0.8721}}} & \textcolor{red}{\textbf{\emph{0}}} & \textcolor{red}{\textbf{\emph{1}}} & \textcolor{red}{\textbf{\emph{1}}} & \textcolor{red}{\textbf{\emph{0}}} & \textcolor{red}{\textbf{\emph{1}}} & \textcolor{red}{\textbf{\emph{1}}} & \textcolor{red}{\textbf{\emph{1}}} \\
\hline \hline
\end{tabular}
\vspace{-0.2em}
\end{table*}

\begin{table*}[!t]
\caption{Metric values of the community structures of \textit{Greedy} $Q$ and \textit{Fine-tuned} $Q$ improved with \textit{Fine-tuned} $Q_{ds}$ on the classical clique network (blue italic font indicates improved score).}
\label{clique_improved}
\vspace{-1.2em}
\centering
\setlength{\tabcolsep}{6pt}
\begin{tabular}{c||c|c|c|c|c|c|c|c|c}
\hline \hline
     ~~Algorithm~~ & $Q$ & $Q_{ds}$ & $VI$ & $NMI$ & $F\textit{-}measure$ &	$NVD$ & $RI$ &	$ARI$ & $JI$ \\
\hline
     Greedy $Q$ improved with Fine-tuned $Q_{ds}$ &  0.8758 & \textcolor{blue}{\textbf{\emph{0.8721}}} & \textcolor{blue}{\textbf{\emph{0}}} & \textcolor{blue}{\textbf{\emph{1}}} & \textcolor{blue}{\textbf{\emph{1}}} & \textcolor{blue}{\textbf{\emph{0}}} & \textcolor{blue}{\textbf{\emph{1}}} & \textcolor{blue}{\textbf{\emph{1}}} & \textcolor{blue}{\textbf{\emph{1}}} \\
\hline
     Fine-tuned $Q$ improved with Fine-tuned $Q_{ds}$ &  0.8758 & \textcolor{blue}{\textbf{\emph{0.8721}}} & \textcolor{blue}{\textbf{\emph{0}}} & \textcolor{blue}{\textbf{\emph{1}}} & \textcolor{blue}{\textbf{\emph{1}}} & \textcolor{blue}{\textbf{\emph{0}}} & \textcolor{blue}{\textbf{\emph{1}}} & \textcolor{blue}{\textbf{\emph{1}}} & \textcolor{blue}{\textbf{\emph{1}}} \\
\hline \hline
\end{tabular}
\vspace{-0.4em}
\end{table*}

\subsubsection{AS Level Internet}
The last real network dataset that is adopted to evaluate the three algorithms is AS level Internet. It is a symmetrized snapshot of the structure of the Internet at the level of autonomous systems, reconstructed from BGP tables posted by the University of Oregon Route Views Project. This snapshot was created by Mark Newman from data for July 22, 2006 and has not been previously published. It has 22963 nodes and 48436 edges.

Table~\ref{as_table} presents the metric values of the community structures detected by the three algorithms. Since this network does not have ground truth communities either, we only calculate $Q$ and $Q_{ds}$. It can be seen from the table that \textit{Fine-tuned} $Q$ and \textit{Fine-tuned} $Q_{ds}$ achieve the highest value of $Q$ and $Q_{ds}$, respectively. Moreover, the $Q_{ds}$ of \textit{Fine-tuned} $Q_{ds}$ is much larger than that of \textit{Greedy} $Q$ and \textit{Fine-tuned} $Q$, which indicates that \textit{Fine-tuned} $Q_{ds}$ performs best on AS level Internet according to $Q_{ds}$, followed by \textit{Fine-tuned} $Q$.

\subsection{Synthetic Networks}
\subsubsection{Clique Network}
\begin{figure}[!t]
\centering
\setlength{\belowcaptionskip}{-1em}
\includegraphics[scale=0.45]{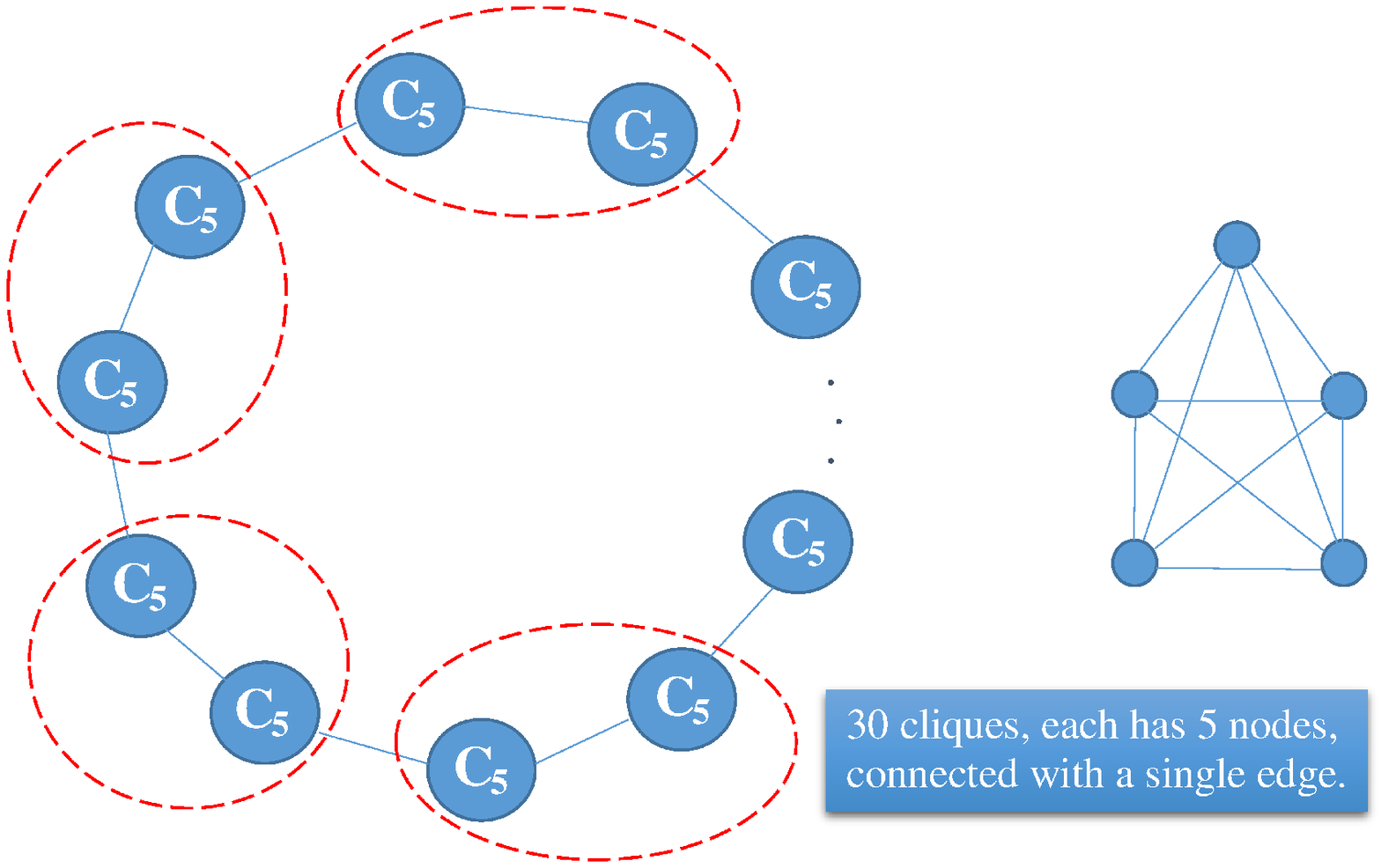}
\vspace{-0.5em}
\centering
\caption{A ring network made out of thirty identical cliques, each having five nodes and connected by single edges.}
\label{30_cliques}
\vspace{-0.5em}
\end{figure}

We now apply the three algorithms to the classical network example \cite{QdsConference,QdsJournal,ResolutionLimit}, displayed in Figure~\ref{30_cliques}, which illustrates modularity ($Q$) has the resolution limit problem. It is a ring network comprised of thirty identical cliques, each of which has five nodes and they are connected by single edges. It is intuitively obvious that each clique forms a single community.

Table~\ref{clique_table} presents the metric values of the community structures detected by the three algorithms. It shows that \textit{Greedy} $Q$ and \textit{Fine-tuned} $Q$ have the same performance. They both achieve the highest value of $Q$ but get about half of the value of $Q_{ds}$ of what \textit{Fine-tuned} $Q_{ds}$ achieves. In fact, \textit{Fine-tuned} $Q_{ds}$ finds exactly thirty communities with each clique being a single community. In contrast, \textit{Greedy} $Q$ and \textit{Fine-tuned} $Q$ discover only sixteen communities with fourteen communities having two cliques and the other two communities having a single clique. Also, we take the community detection results of \textit{Greedy} $Q$ and \textit{Fine-tuned} $Q$ as the input to \textit{Fine-tuned} $Q_{ds}$ to try to improve those results. The metric values of the community structures after improvement with \textit{Fine-tuned} $Q_{ds}$ are recorded in Table~\ref{clique_improved}. This table shows that the community structures discovered are identical to that of \textit{Fine-tuned} $Q_{ds}$, which means that the results of \textit{Greedy} $Q$ and \textit{Fine-tuned} $Q$ are dramatically improved with \textit{Fine-tuned} $Q_{ds}$. Therefore, it can be concluded from Tables~\ref{clique_table} and~\ref{clique_improved} that a larger value of $Q_{ds}$ (but not $Q$) implies a higher quality of the community structure. Moreover, $Q_{ds}$ solves the resolution limit problem of $Q$. Finally, \textit{Fine-tuned} $Q_{ds}$ is effective in maximizing $Q_{ds}$ and in finding meaningful community structure.

\begin{table*}[!t]
\caption{Metric values of the community structures of \textit{Greedy} $Q$ on the LFR benchmark networks with $(\gamma, \beta)=(2, 1)$.} 
\label{lfr_greedy_Q}
\vspace{-1.2em}
\centering
\setlength{\tabcolsep}{10pt}
\begin{tabular}{c||c|c|c|c|c|c|c|c|c}
\hline \hline
     $~~~~\mu~~~~$ & $Q$ & $Q_{ds}$ & $VI$ & $NMI$ & $F\textit{-}measure$ &	$NVD$ & $RI$ &	$ARI$ & $JI$ \\
\hline
     0.05 & \textcolor{red}{\textbf{\emph{0.9021}}} & \textcolor{red}{\textbf{\emph{0.4481}}} & \textcolor{red}{\textbf{\emph{0.2403}}} & \textcolor{red}{\textbf{\emph{0.9767}}} & \textcolor{red}{\textbf{\emph{0.9382}}} & \textcolor{red}{\textbf{\emph{0.0399}}} & \textcolor{red}{\textbf{\emph{0.9959}}} & \textcolor{red}{\textbf{\emph{0.9308}}} & \textcolor{red}{\textbf{\emph{0.8758}}} \\
\hline
     0.1 & \textcolor{red}{\textbf{\emph{0.8461}}} & 0.3546 & 0.5213 & \textcolor{red}{\textbf{\emph{0.9482}}} & 0.8539 & 0.0912 & 0.9882 & 0.821 & 0.7089\\
\hline
     0.15 & 0.7862 & 0.2604 & 0.8537 & 0.9125 & 0.7604 & 0.1432 & 0.9776 & 0.7042 & 0.5573\\
\hline
     0.2 & \textcolor{red}{\textbf{\emph{0.7256}}} & 0.1934 & 1.3601 & 0.8579 & 0.6314 & 0.2173 & 0.9601 & 0.5445 & 0.3911\\
\hline
     0.25 & 0.6612 & 0.1411 & 1.7713 & 0.8093 & 0.5477 & 0.2642 & 0.9444 & 0.4498 & 0.309\\
\hline
     0.3 & 0.5959 & 0.09377 & 2.1758 & 0.7493 & 0.4745 & 0.3085 & 0.921 & 0.3779 & 0.255\\
\hline
     0.35 & 0.545 & 0.07237 & 2.4599 & 0.7122 & 0.4182 & 0.3347 & 0.9045 & 0.3206 & 0.2134\\
\hline
     0.4 & 0.4857 & 0.05521 & 2.7444 & 0.672 & 0.3745 & 0.3623 & 0.8874 & 0.2766 & 0.1836\\
\hline
     0.45 & 0.4356 & 0.04133 & 3.0108 & 0.6289 & 0.327 & 0.3875 & 0.8617 & 0.2288 & 0.153\\
\hline
     0.5 & \textcolor{red}{\textbf{\emph{0.3803}}} & 0.03016 & 3.4296 & 0.5685 & 0.2874 & 0.4159 & 0.8386 & 0.1885 & 0.1282\\
\hline \hline
\end{tabular}
\vspace{-0.2em}
\end{table*}

\begin{table*}[!t]
\caption{Metric values of the community structures of \textit{Fine-tuned} $Q$ on the LFR benchmark networks with $(\gamma, \beta)=(2, 1)$.} 
\label{lfr_fine_tuned_Q}
\vspace{-1.2em}
\centering
\setlength{\tabcolsep}{10pt}
\begin{tabular}{c||c|c|c|c|c|c|c|c|c}
\hline \hline
     $~~~~\mu~~~~$ & $Q$ & $Q_{ds}$ & $VI$ & $NMI$ & $F\textit{-}measure$ &	$NVD$ & $RI$ &	$ARI$ & $JI$ \\
\hline
     0.05 & 0.8411 & 0.3875 & 0.8674 & 0.8868 & 0.8137 & 0.1049 & 0.9404 & 0.7503 & 0.6673\\
\hline
     0.1 & 0.8419 & 0.3837 & \textcolor{red}{\textbf{\emph{0.5195}}} & 0.9481 & 0.8851 & 0.0695 & 0.9875 & 0.8333 & 0.7408\\
\hline
     0.15 & \textcolor{red}{\textbf{\emph{0.7886}}} & 0.3324 & \textcolor{red}{\textbf{\emph{0.6453}}} & 0.9358 & 0.8664 & 0.0844 & 0.9858 & 0.801 & 0.6921\\
\hline
     0.2 & 0.7221 & 0.2922 & 0.9615 & 0.9022 & 0.8056 & 0.1222 & 0.9725 & 0.7099 & 0.6061\\
\hline
     0.25 & \textcolor{red}{\textbf{\emph{0.6694}}} & 0.2502 & 1.11 & 0.8833 & 0.7831 & 0.137 & 0.9594 & 0.7045 & 0.5939\\
\hline
     0.3 & \textcolor{red}{\textbf{\emph{0.626}}} & 0.2022 & \textcolor{red}{\textbf{\emph{1.0722}}} & \textcolor{red}{\textbf{\emph{0.892}}} & 0.813 & 0.1265 & 0.9811 & 0.7317 & 0.5963\\
\hline
     0.35 & \textcolor{red}{\textbf{\emph{0.5479}}} & 0.1516 & 1.6786 & 0.8153 & 0.705 & 0.1942 & 0.949 & 0.5963 & 0.4629\\
\hline
     0.4 & 0.5044 & 0.124 & 1.8382 & 0.8108 & 0.6935 & 0.2111 & 0.9646 & 0.5592 & 0.4118\\
\hline
     0.45 & 0.4274 & 0.07865 & 2.5657 & 0.7274 & 0.5913 & 0.2863 & 0.9463 & 0.4419 & 0.3129\\
\hline
     0.5 & 0.3766 & 0.05808 & 3.0333 & 0.675 & 0.5328 & 0.3375 & \textcolor{red}{\textbf{\emph{0.9366}}} & 0.3721 & 0.2537\\
\hline \hline
\end{tabular}
\vspace{-0.2em}
\end{table*}

\begin{table*}[!t]
\caption{Metric values of the community structures of \textit{Fine-tuned} $Q_{ds}$ on the LFR benchmark networks with $(\gamma, \beta)=(2, 1)$.} 
\label{lfr_fine_tuned_Qds}
\vspace{-1.2em}
\centering
\setlength{\tabcolsep}{10pt}
\begin{tabular}{c||c|c|c|c|c|c|c|c|c}
\hline \hline
     $~~~~\mu~~~~$ & $Q$ & $Q_{ds}$ & $VI$ & $NMI$ & $F\textit{-}measure$ &	$NVD$ & $RI$ &	$ARI$ & $JI$ \\
\hline
     0.05 & 0.845 & 0.4257 & 0.8112 & 0.9186 & 0.8564 & 0.09585 & 0.9736 & 0.691 & 0.5717\\
\hline
     0.1 & 0.7934 & \textcolor{red}{\textbf{\emph{0.4144}}} & 0.5809 & 0.9447 & \textcolor{red}{\textbf{\emph{0.9326}}} & \textcolor{red}{\textbf{\emph{0.0625}}} & \textcolor{red}{\textbf{\emph{0.9915}}} & \textcolor{red}{\textbf{\emph{0.8566}}} & \textcolor{red}{\textbf{\emph{0.7646}}} \\
\hline
     0.15 & 0.7426 & \textcolor{red}{\textbf{\emph{0.3605}}} & 0.6769 & \textcolor{red}{\textbf{\emph{0.9359}}} & \textcolor{red}{\textbf{\emph{0.9172}}} & \textcolor{red}{\textbf{\emph{0.0711}}} & \textcolor{red}{\textbf{\emph{0.9902}}} & \textcolor{red}{\textbf{\emph{0.8303}}} & \textcolor{red}{\textbf{\emph{0.7225}}} \\
\hline
     0.2 & 0.6786 & \textcolor{red}{\textbf{\emph{0.337}}} & \textcolor{red}{\textbf{\emph{0.7824}}} & \textcolor{red}{\textbf{\emph{0.9278}}} & \textcolor{red}{\textbf{\emph{0.9195}}} & \textcolor{red}{\textbf{\emph{0.0795}}} & \textcolor{red}{\textbf{\emph{0.9908}}} & \textcolor{red}{\textbf{\emph{0.8186}}} & \textcolor{red}{\textbf{\emph{0.7037}}} \\
\hline
     0.25 & 0.6202 & \textcolor{red}{\textbf{\emph{0.2891}}} & \textcolor{red}{\textbf{\emph{1.0244}}} & \textcolor{red}{\textbf{\emph{0.9046}}} & \textcolor{red}{\textbf{\emph{0.8909}}} & \textcolor{red}{\textbf{\emph{0.106}}} & \textcolor{red}{\textbf{\emph{0.9868}}} & \textcolor{red}{\textbf{\emph{0.7575}}} & \textcolor{red}{\textbf{\emph{0.6253}}} \\
\hline
     0.3 & 0.5693 & \textcolor{red}{\textbf{\emph{0.235}}} & 1.1347 & 0.8919 & \textcolor{red}{\textbf{\emph{0.8874}}} & \textcolor{red}{\textbf{\emph{0.1183}}} & \textcolor{red}{\textbf{\emph{0.9845}}} & \textcolor{red}{\textbf{\emph{0.7372}}} & \textcolor{red}{\textbf{\emph{0.5983}}} \\
\hline
     0.35 & 0.5443 & \textcolor{red}{\textbf{\emph{0.2244}}} & \textcolor{red}{\textbf{\emph{0.9401}}} & \textcolor{red}{\textbf{\emph{0.9123}}} & \textcolor{red}{\textbf{\emph{0.9129}}} & \textcolor{red}{\textbf{\emph{0.09585}}} & \textcolor{red}{\textbf{\emph{0.989}}} & \textcolor{red}{\textbf{\emph{0.7984}}} & \textcolor{red}{\textbf{\emph{0.6783}}} \\
\hline
     0.4 & \textcolor{red}{\textbf{\emph{0.505}}} & \textcolor{red}{\textbf{\emph{0.1964}}} & \textcolor{red}{\textbf{\emph{0.9444}}} & \textcolor{red}{\textbf{\emph{0.9123}}} & \textcolor{red}{\textbf{\emph{0.9091}}} & \textcolor{red}{\textbf{\emph{0.0966}}} & \textcolor{red}{\textbf{\emph{0.989}}} & \textcolor{red}{\textbf{\emph{0.7929}}} & \textcolor{red}{\textbf{\emph{0.668}}} \\
\hline
     0.45 & \textcolor{red}{\textbf{\emph{0.4536}}} & \textcolor{red}{\textbf{\emph{0.1632}}} & \textcolor{red}{\textbf{\emph{1.1523}}} & \textcolor{red}{\textbf{\emph{0.8925}}} & \textcolor{red}{\textbf{\emph{0.8806}}} & \textcolor{red}{\textbf{\emph{0.1196}}} & \textcolor{red}{\textbf{\emph{0.9834}}} & \textcolor{red}{\textbf{\emph{0.7337}}} & \textcolor{red}{\textbf{\emph{0.6021}}} \\
\hline
     0.5 & 0.3563 & \textcolor{red}{\textbf{\emph{0.1196}}} & \textcolor{red}{\textbf{\emph{1.9677}}} & \textcolor{red}{\textbf{\emph{0.8036}}} & \textcolor{red}{\textbf{\emph{0.7489}}} & \textcolor{red}{\textbf{\emph{0.2076}}} & 0.9213 & \textcolor{red}{\textbf{\emph{0.4984}}} & \textcolor{red}{\textbf{\emph{0.3813}}} \\
\hline \hline
\end{tabular}
\vspace{-0.5em}
\end{table*}

\subsubsection{LFR Benchmark Networks}
To further compare the performance of \textit{Greedy} $Q$, \textit{Fine-tuned} $Q$, and \textit{Fine-tuned} $Q_{ds}$, we choose the LFR benchmark networks  \cite{LFR} which have become a standard in the evaluation of the performance of community detection algorithms and also have known ground truth communities. The LFR benchmark network that we used here has $1000$ nodes with average degree $15$ and maximum degree $50$. The exponent $\gamma$ for the degree sequence varies from $2$ to $3$. The exponent $\beta$ for the community size distribution ranges from $1$ to $2$. Then, four pairs of the exponents $(\gamma, \beta) = (2, 1), (2, 2), (3, 1), \text{and}~(3, 2)$ are chosen in order to explore the widest spectrum of graph structures. The mixing parameter $\mu$ is varied from $0.05$ to $0.5$. It means that each node shares a fraction $(1- \mu)$ of its edges with the other nodes in its community and shares a fraction $\mu$ of its edges with the nodes outside its community. Thus, low mixing parameters indicate strong community structure. Also, we generate $10$ network instances for each $\mu$. Hence, each metric value in Tables~\ref{lfr_greedy_Q}-\ref{lfr_fine_tuned_Q_improved} represents the average metric values of all $10$ instances. Since the experimental results are similar for all four pairs of exponents $(\gamma, \beta) = (2, 1), (2, 2), (3, 1), \text{and}~(3, 2)$, for the sake of brevity, we only present the results for $(\gamma, \beta) = (2, 1)$ here.

Tables~\ref{lfr_greedy_Q}-\ref{lfr_fine_tuned_Qds} show the metric values of the community structures detected with \textit{Greedy} $Q$, \textit{Fine-tuned} $Q$, and \textit{Fine-tuned} $Q_{ds}$, respectively, on the LFR benchmark networks with $(\gamma, \beta)=(2, 1)$ and $\mu$ varying from $0.05$ to $0.5$. The red italic font in the table denotes that the corresponding algorithm achieves the best value for a certain quality metric among the three algorithms. The results in these tables show that \textit{Greedy} $Q$ obtains the best values for all the nine measurements when $\mu=0.05$, while \textit{Fine-tuned} $Q_{ds}$ achieves the highest values of $Q_{ds}$ and the best values for almost all the seven metrics based on ground truth communities when $\mu$ ranges from $0.1$ to $0.5$. Also, \textit{Fine-tuned} $Q$ gets the second best values for $Q_{ds}$ and almost all the seven metrics in the same range of $\mu$. However, for $Q$ the best is \textit{Greedy} $Q$, followed by \textit{Fine-tuned} $Q$, and \textit{Fine-tuned} $Q_{ds}$ is the last.

\begin{table*}[!t]
\caption{Metric values of the community structures of \textit{Greedy} $Q$ improved with \textit{Fine-tuned} $Q_{ds}$ on the LFR benchmark networks with $(\gamma, \beta)=(2, 1)$.} 
\label{lfr_greedy_Q_improved}
\vspace{-1.2em}
\centering
\setlength{\tabcolsep}{10pt}
\begin{tabular}{c||c|c|c|c|c|c|c|c|c}
\hline \hline
     $~~~~\mu~~~~$ & $Q$ & $Q_{ds}$ & $VI$ & $NMI$ & $F\textit{-}measure$ &	$NVD$ & $RI$ &	$ARI$ & $JI$ \\
\hline
     0.05 & 0.8743 & \textcolor{blue}{\textbf{\emph{0.4979}}} & \textcolor{blue}{\textbf{\emph{0.2131}}} & \textcolor{blue}{\textbf{\emph{0.98}}} & \textcolor{blue}{\textbf{\emph{0.9784}}} & \textcolor{blue}{\textbf{\emph{0.02195}}} & \textcolor{blue}{\textbf{\emph{0.997}}} & \textcolor{blue}{\textbf{\emph{0.943}}} & \textcolor{blue}{\textbf{\emph{0.895}}}\\
\hline
     0.1 & 0.8246 & \textcolor{blue}{\textbf{\emph{0.4522}}} & \textcolor{blue}{\textbf{\emph{0.2428}}} & \textcolor{blue}{\textbf{\emph{0.9773}}} & \textcolor{blue}{\textbf{\emph{0.9762}}} & \textcolor{blue}{\textbf{\emph{0.02395}}} & \textcolor{blue}{\textbf{\emph{0.9967}}} & \textcolor{blue}{\textbf{\emph{0.9379}}} & \textcolor{blue}{\textbf{\emph{0.8864}}}\\
\hline
     0.15 & 0.7716 & \textcolor{blue}{\textbf{\emph{0.4013}}} & \textcolor{blue}{\textbf{\emph{0.2972}}} & \textcolor{blue}{\textbf{\emph{0.9722}}} & \textcolor{blue}{\textbf{\emph{0.9719}}} & \textcolor{blue}{\textbf{\emph{0.0289}}} & \textcolor{blue}{\textbf{\emph{0.9962}}} & \textcolor{blue}{\textbf{\emph{0.9269}}} & \textcolor{blue}{\textbf{\emph{0.8674}}}\\
\hline
     0.2 & 0.7232 & \textcolor{blue}{\textbf{\emph{0.384}}} & \textcolor{blue}{\textbf{\emph{0.3503}}} & \textcolor{blue}{\textbf{\emph{0.9679}}} & \textcolor{blue}{\textbf{\emph{0.9664}}} & \textcolor{blue}{\textbf{\emph{0.03505}}} & \textcolor{blue}{\textbf{\emph{0.9959}}} & \textcolor{blue}{\textbf{\emph{0.9163}}} & \textcolor{blue}{\textbf{\emph{0.8496}}} \\
\hline
     0.25 & \textcolor{blue}{\textbf{\emph{0.6667}}} & \textcolor{blue}{\textbf{\emph{0.3347}}} & \textcolor{blue}{\textbf{\emph{0.4474}}} & \textcolor{blue}{\textbf{\emph{0.9592}}} & \textcolor{blue}{\textbf{\emph{0.9582}}} & \textcolor{blue}{\textbf{\emph{0.04485}}} & \textcolor{blue}{\textbf{\emph{0.9953}}} & \textcolor{blue}{\textbf{\emph{0.9011}}} & \textcolor{blue}{\textbf{\emph{0.8243}}} \\
\hline
     0.3 & \textcolor{blue}{\textbf{\emph{0.6094}}} & \textcolor{blue}{\textbf{\emph{0.2619}}} & \textcolor{blue}{\textbf{\emph{0.6061}}} & \textcolor{blue}{\textbf{\emph{0.9432}}} & \textcolor{blue}{\textbf{\emph{0.9457}}} & \textcolor{blue}{\textbf{\emph{0.05905}}} & \textcolor{blue}{\textbf{\emph{0.9934}}} & \textcolor{blue}{\textbf{\emph{0.876}}} & \textcolor{blue}{\textbf{\emph{0.7856}}} \\
\hline
     0.35 & \textcolor{blue}{\textbf{\emph{0.5584}}} & \textcolor{blue}{\textbf{\emph{0.2377}}} & \textcolor{blue}{\textbf{\emph{0.691}}} & \textcolor{blue}{\textbf{\emph{0.9364}}} & \textcolor{blue}{\textbf{\emph{0.94}}} & \textcolor{blue}{\textbf{\emph{0.0697}}} & \textcolor{blue}{\textbf{\emph{0.9931}}} & \textcolor{blue}{\textbf{\emph{0.8615}}} & \textcolor{blue}{\textbf{\emph{0.7626}}} \\
\hline
     0.4 & \textcolor{blue}{\textbf{\emph{0.5062}}} & \textcolor{blue}{\textbf{\emph{0.199}}} & \textcolor{blue}{\textbf{\emph{0.8285}}} & \textcolor{blue}{\textbf{\emph{0.9236}}} & \textcolor{blue}{\textbf{\emph{0.9247}}} & \textcolor{blue}{\textbf{\emph{0.0823}}} & \textcolor{blue}{\textbf{\emph{0.9916}}} & \textcolor{blue}{\textbf{\emph{0.8376}}} & \textcolor{blue}{\textbf{\emph{0.7281}}} \\
\hline
     0.45 & \textcolor{blue}{\textbf{\emph{0.4587}}} & \textcolor{blue}{\textbf{\emph{0.169}}} & \textcolor{blue}{\textbf{\emph{0.9016}}} & \textcolor{blue}{\textbf{\emph{0.9172}}} & \textcolor{blue}{\textbf{\emph{0.9222}}} & \textcolor{blue}{\textbf{\emph{0.0904}}} & \textcolor{blue}{\textbf{\emph{0.9914}}} & \textcolor{blue}{\textbf{\emph{0.8252}}} & \textcolor{blue}{\textbf{\emph{0.7099}}} \\
\hline
     0.5 & \textcolor{blue}{\textbf{\emph{0.4014}}} & \textcolor{blue}{\textbf{\emph{0.1385}}} & \textcolor{blue}{\textbf{\emph{1.2004}}} & \textcolor{blue}{\textbf{\emph{0.8906}}} & \textcolor{blue}{\textbf{\emph{0.8938}}} & \textcolor{blue}{\textbf{\emph{0.1215}}} & \textcolor{blue}{\textbf{\emph{0.9885}}} & \textcolor{blue}{\textbf{\emph{0.7686}}} & \textcolor{blue}{\textbf{\emph{0.6326}}} \\
\hline \hline
\end{tabular}
\vspace{-0.2em}
\end{table*}

\begin{table*}[!t]
\caption{Metric values of the community structures of \textit{Fine-tuned} $Q$ improved with \textit{Fine-tuned} $Q_{ds}$ on the LFR benchmark networks with $(\gamma, \beta)=(2, 1)$.} 
\label{lfr_fine_tuned_Q_improved}
\vspace{-1.2em}
\centering
\setlength{\tabcolsep}{10pt}
\begin{tabular}{c||c|c|c|c|c|c|c|c|c}
\hline \hline
     $~~~~\mu~~~~$ & $Q$ & $Q_{ds}$ & $VI$ & $NMI$ & $F\textit{-}measure$ &	$NVD$ & $RI$ &	$ARI$ & $JI$ \\
\hline
     0.05 & \textcolor{blue}{\textbf{\emph{0.8519}}} & \textcolor{blue}{\textbf{\emph{0.4463}}} & \textcolor{blue}{\textbf{\emph{0.5949}}} & \textcolor{blue}{\textbf{\emph{0.937}}} & \textcolor{blue}{\textbf{\emph{0.8954}}} & \textcolor{blue}{\textbf{\emph{0.0709}}} & \textcolor{blue}{\textbf{\emph{0.9781}}} & \textcolor{blue}{\textbf{\emph{0.8177}}} & \textcolor{blue}{\textbf{\emph{0.7377}}} \\
\hline
     0.1 & 0.8186 & \textcolor{blue}{\textbf{\emph{0.4397}}} & \textcolor{blue}{\textbf{\emph{0.3405}}} & \textcolor{blue}{\textbf{\emph{0.9679}}} & \textcolor{blue}{\textbf{\emph{0.9615}}} & \textcolor{blue}{\textbf{\emph{0.03415}}} & \textcolor{blue}{\textbf{\emph{0.9952}}} & \textcolor{blue}{\textbf{\emph{0.9125}}} & \textcolor{blue}{\textbf{\emph{0.8452}}} \\
\hline
     0.15 & 0.769 & \textcolor{blue}{\textbf{\emph{0.391}}} & \textcolor{blue}{\textbf{\emph{0.4285}}} & \textcolor{blue}{\textbf{\emph{0.9597}}} & \textcolor{blue}{\textbf{\emph{0.9533}}} & \textcolor{blue}{\textbf{\emph{0.0432}}} & \textcolor{blue}{\textbf{\emph{0.9946}}} & \textcolor{blue}{\textbf{\emph{0.8993}}} & \textcolor{blue}{\textbf{\emph{0.8231}}} \\
\hline
     0.2 & 0.7185 & \textcolor{blue}{\textbf{\emph{0.369}}} & \textcolor{blue}{\textbf{\emph{0.4654}}} & \textcolor{blue}{\textbf{\emph{0.9571}}} & \textcolor{blue}{\textbf{\emph{0.9479}}} & \textcolor{blue}{\textbf{\emph{0.04975}}} & \textcolor{blue}{\textbf{\emph{0.9943}}} & \textcolor{blue}{\textbf{\emph{0.8853}}} & \textcolor{blue}{\textbf{\emph{0.8014}}} \\
\hline
     0.25 & 0.6672 & \textcolor{blue}{\textbf{\emph{0.326}}} & \textcolor{blue}{\textbf{\emph{0.5667}}} & \textcolor{blue}{\textbf{\emph{0.9477}}} & \textcolor{blue}{\textbf{\emph{0.9365}}} & \textcolor{blue}{\textbf{\emph{0.05805}}} & \textcolor{blue}{\textbf{\emph{0.9936}}} & \textcolor{blue}{\textbf{\emph{0.8713}}} & \textcolor{blue}{\textbf{\emph{0.7785}}} \\
\hline
     0.3 & 0.6109 & \textcolor{blue}{\textbf{\emph{0.2598}}} & \textcolor{blue}{\textbf{\emph{0.6962}}} & \textcolor{blue}{\textbf{\emph{0.9346}}} & \textcolor{blue}{\textbf{\emph{0.9372}}} & \textcolor{blue}{\textbf{\emph{0.06505}}} & \textcolor{blue}{\textbf{\emph{0.9926}}} & \textcolor{blue}{\textbf{\emph{0.8609}}} & \textcolor{blue}{\textbf{\emph{0.762}}} \\
\hline
     0.35 & 0.5474 & \textcolor{blue}{\textbf{\emph{0.2297}}} & \textcolor{blue}{\textbf{\emph{0.9525}}} & \textcolor{blue}{\textbf{\emph{0.9108}}} & \textcolor{blue}{\textbf{\emph{0.9175}}} & \textcolor{blue}{\textbf{\emph{0.0961}}} & \textcolor{blue}{\textbf{\emph{0.9882}}} & \textcolor{blue}{\textbf{\emph{0.7963}}} & \textcolor{blue}{\textbf{\emph{0.6821}}} \\
\hline
     0.4 & 0.4966 & \textcolor{blue}{\textbf{\emph{0.1983}}} & \textcolor{blue}{\textbf{\emph{1.0601}}} & \textcolor{blue}{\textbf{\emph{0.9021}}} & \textcolor{blue}{\textbf{\emph{0.9118}}} & \textcolor{blue}{\textbf{\emph{0.1029}}} & \textcolor{blue}{\textbf{\emph{0.9896}}} & \textcolor{blue}{\textbf{\emph{0.7963}}} & \textcolor{blue}{\textbf{\emph{0.672}}} \\
\hline
     0.45 & \textcolor{blue}{\textbf{\emph{0.4284}}} & \textcolor{blue}{\textbf{\emph{0.1535}}} & \textcolor{blue}{\textbf{\emph{1.4754}}} & \textcolor{blue}{\textbf{\emph{0.8635}}} & \textcolor{blue}{\textbf{\emph{0.8694}}} & \textcolor{blue}{\textbf{\emph{0.1486}}} & \textcolor{blue}{\textbf{\emph{0.9831}}} & \textcolor{blue}{\textbf{\emph{0.6836}}} & \textcolor{blue}{\textbf{\emph{0.5362}}} \\
\hline
     0.5 & 0.3654 & \textcolor{blue}{\textbf{\emph{0.1258}}} & \textcolor{blue}{\textbf{\emph{1.9271}}} & \textcolor{blue}{\textbf{\emph{0.8192}}} & \textcolor{blue}{\textbf{\emph{0.8193}}} & \textcolor{blue}{\textbf{\emph{0.1987}}} & \textcolor{blue}{\textbf{\emph{0.968}}} & \textcolor{blue}{\textbf{\emph{0.5852}}} & \textcolor{blue}{\textbf{\emph{0.4423}}} \\
\hline \hline
\end{tabular}
\vspace{-0.5em}
\end{table*}

In summary, the seven measurements based on ground truth communities are all consistent with $Q_{ds}$, but not consistent with $Q$. This consistency indicates the superiority of $Q_{ds}$ over $Q$ as a community quality metric. In addition, \textit{Fine-tuned} $Q_{ds}$ performs best among the three algorithms for $\mu>0.05$, which demonstrates that it is very effective and does a very good job in optimizing $Q_{ds}$.

We then take the community detection results of \textit{Greedy} $Q$ and \textit{Fine-tuned} $Q$ as the input to \textit{Fine-tuned} $Q_{ds}$ to improve those results. The measurement values of the community structures after improvement with \textit{Fine-tuned} $Q_{ds}$ are displayed in Tables~\ref{lfr_greedy_Q_improved} and \ref{lfr_fine_tuned_Q_improved}. The blue italic font in Table~\ref{lfr_greedy_Q_improved} and Table~\ref{lfr_fine_tuned_Q_improved} implies that the metric value in these two tables is improved compared to the one in Table~\ref{lfr_greedy_Q} and that in Table~\ref{lfr_fine_tuned_Q}, respectively. Then, compared with those of \textit{Greedy} $Q$ shown in Table~\ref{lfr_greedy_Q} and those of \textit{Fine-tuned} $Q$ shown in Table~\ref{lfr_fine_tuned_Q}, all measurements, except in some cases for $Q$, are significantly improved with \textit{Fine-tuned} $Q_{ds}$. This again indicates that all the seven metrics described in Subsection~\ref{evaluation_metrics} are consistent with $Q_{ds}$, but not consistent with $Q$. Interestingly, those results are even better than those of \textit{Fine-tuned} $Q_{ds}$ itself presented in Table~\ref{lfr_fine_tuned_Qds}. Thus, it can be concluded that \textit{Fine-tuned} $Q_{ds}$ is very powerful in improving the community detection results of other algorithms.

\section{Conclusion}
In this paper, we review the definition of modularity and its corresponding maximization methods. Moreover, we show that modularity optimization has two opposite but coexisting issues. We also review several community quality metrics proposed to solve the resolution limit problem. We then discuss our \textit{Modularity Density} ($Q_{ds}$) metric which simultaneously avoids those two problems. Finally, we propose an efficient and effective fine-tuned algorithm to maximize $Q_{ds}$. This new algorithm can actually be used to optimize any community quality metric. We evaluate the three algorithms, \textit{Greedy} $Q$, \textit{Fine-tuned} $Q$ based on $Q$, and \textit{Fine-tuned} $Q_{ds}$ based on $Q_{ds}$, with seven metrics based on ground truth communities. These evaluations are done on four real networks, and also on the classical clique network and the LFR benchmark networks, each instance of the last is defined with parameters selected from wide range of their values. The results demonstrate that \textit{Fine-tuned} $Q_{ds}$ performs best among the three algorithms, followed by \textit{Fine-tuned} $Q$. The experiments also show that \textit{Fine-tuned} $Q_{ds}$ can dramatically improve the community detection results of other algorithms. In addition, all the seven quality metrics based on ground truth communities are consistent with $Q_{ds}$, but not consistent with $Q$, which indicates the superiority of $Q_{ds}$ over $Q$ as a community quality metric.


%



\ifCLASSOPTIONcompsoc
  \section*{Acknowledgments}
\else
  \section*{Acknowledgment}
\fi
This work was supported in part by the Army Research Laboratory under Cooperative Agreement Number W911NF-09-2-0053 and by the the Office of Naval Research Grant No. N00014-09-1-0607. The views and conclusions contained in this document are those of the authors and should not be interpreted as representing the official policies either expressed or implied of the Army Research Laboratory or the U.S. Government.


\ifCLASSOPTIONcaptionsoff
  \newpage
\fi



%

%

\bibliographystyle{TCSS}
\bibliography{TCSS}

%








\end{document}